\documentclass[useAMS,usenatbib,]{mn2e}
\usepackage{times}
\usepackage{natbib}
\usepackage[dvips]{graphicx}
\usepackage{graphicx}
\usepackage{amsmath}
\usepackage{txfonts}
\usepackage{natbib}
\usepackage{threeparttable}
 
\title[] {Discovery of radio halos and double-relics in distant MACS
  galaxy clusters: clues to the efficiency of particle
  acceleration.}

\author[A. Bonafede et al.]
{A. Bonafede$^{1,2}$\thanks{E-mail: a.bonafede@jacobs-university.de}, M. Br\"uggen$^{1,2}$, R. van Weeren$^{3,4}$, F. Vazza$^{1,2}$, G. Giovannini$^{5,6}$
 \newauthor H. Ebeling$^{7}$, A. C. Edge$^{8}$, M. Hoeft$^9$, U. Klein$^{10}$\\
$^1$Jacobs University Bremen, Campus Ring 1, D-28759 Bremen,  Germany.\\ 
$^2$ Hamburger Sternwarte, Universit\"at Hamburg, Gojenbergsweg 112, 21029, Hamburg, Germany. \\
$^3$ Netherlands Institute for Radio Astronomy (ASTRON), Post-bus 2, 7990 AA Dwingeloo, the Netherlands. \\
$^4$ Leiden Observatory, Leiden University, P.O. Box 9513, NL-2300 RA Leiden, The Netherlands. \\
$^5$ INAF Istituto di Radioastronomia, via P. Gobetti 101, I-40129 Bologna, Italy. \\
$^6$ Universit\`a di Bologna, Dip. di Astronomia, via Ranzani 1, I-40126 Bologna, Italy. \\
$^7$ Institute for Astronomy, University of Hawaii, 2680 Woodlawn Drive, Honolulu, HI 96822, USA.\\
$^8$ Department of Physics, Durham University, Durham DH1 3LE, United Kingdom. \\
$^9$ Th\"uringer Landessternwarte, Sternwarte 5, 07778 Tautemburg, Germany.\\
$^{10}$ Argelander-Institut f\"ur Astronomie, Auf dem H\"ugel 71, D-53121 Bonn, Germany.
}

\begin{document}  
   \date{Received ; accepted } 

\maketitle
\begin{abstract}  
  We have performed 323 MHz observations with the Giant Metrewave
  Radio Telescope of the most promising candidates selected from the
  MACS catalog. The aim of the work is to extend our knowledge of the
  radio halo and relic populations to $z>0.3$, the epoch in
  which massive clusters formed. In MACSJ1149.5+2223 and
  MACSJ1752.1+4440, we discovered two double-relic systems
  with a radio halo, and in MACSJ0553.4-3342 we found
  a radio halo. Archival Very Large Array observations and
  Westerbork Synthesis Radio Telescope observations have been used to
  study the polarization and spectral index properties. The radio
  halo in MACSJ1149.5+2223 has the steepest spectrum ever found so far
  in these objects ($\alpha \geq$ 2).  The double relics in
  MACSJ1149.5+2223 are peculiar in their position that is misaligned
  with the main merger axis. The relics are polarized up to 30\% and
  40\% in MACSJ1149.5+2223 and MACSJ1752.040+44, respectively. In both cases, the
  magnetic field is roughly aligned with the relics' main axes. The
  spectra in the relics in MACSJ1752.040+44 steepen
  towards the cluster centre, in agreement with model expectations.
  X-ray data on MACSJ0553.4-3342 suggests that this cluster is
  undergoing a major merger, with the merger axis close to the plane
  of the sky. The cores of the disrupted clusters have just passed
  each other, but no radio relic is detected in this system. If
  turbulence is responsible for the radio emission, we argue that it
  must develop before the core passage. A comparison of double relic
  plus halo system with cosmological simulations allows a simultaneous
  estimate of the acceleration efficiencies at shocks (to produce
  relics) and of turbulence (to produce the halo). 
\end{abstract}

%

\section{Introduction}

A fraction of galaxy clusters host diffuse radio emission, that is not
connected to any of the cluster radiogalaxies. These radio sources are
classified as radio halos and radio relics, depending on their
location and morphology. In all these sources relativistic particles
need to be (re)accelerated in order to produce the observed radio
emission, even though the underlying physical mechanisms are likely to
be different. Radio halos permeate the central Mpc of galaxy clusters,
and in some cases the radio emission roughly follows the X-ray
emission from the thermal gas \citep[see review by][and references
  therein]{2008SSRv..134...93F}. Radio halos are characterized by a
steep radio spectrum,\footnote{We define here the spectrum as $S(\nu)
  \propto \nu^{-\alpha}$. } with $\alpha\geq 1.2$, and their power at
1.4 GHz correlates with the X-ray luminosity of the host cluster
\citep{2000ApJ...544..686L,Giovannini09}. The origin of radio halos is
still unknown, although a clear connection between the presence of a
radio halo and the merging state of the host cluster is present
\citep{2001ApJ...553L..15B,2010ApJ...721L..82C,2011ApJ...740L..28B,2012MNRAS.421L.112B}.
The models proposed so far can be divided into two classes: ``hadronic
models''
\citep[e.g][]{1980ApJ...239L..93D,2010ApJ...722..737K,2011A&A...527A..99E}
and ``re-acceleration models''. It was recently shown that hadronic
models fail to reproduce the observed radio emission of the Coma
cluster, once the upper limits by FERMI and the magnetic field
estimate from Faraday Rotation Measures are combined
\citep{2010MNRAS.401...47D}. Recent LOFAR radio observations of the
cluster Abell 2256 have shown that the spectral flattening predicted
at low frequencies by re-acceleration models are not observed \citep{2012arXiv1205.4730V}.  Hence, the question about the origin
of radio halos is still open, and it is likely that more complex
scenarios have to be considered.\\ Radio relics are irregularly shaped
radio sources, located at the outskirts of galaxy clusters. They
usually have an arc-like structure, and are found to be polarized at
10 - 80\% level \citep[see review by][and references
  therein]{2011SSRv..tmp..138B}. Like radio halos, radio relics are
characterized by a steep spectrum and a weak surface brightness at 20
cm that makes their detection difficult. Their origin is still
unclear, but there is a common consensus that they are related to
merger shocks.  A merging shock could accelerate particles via
Diffusive Shock Acceleration (DSA), and amplify the magnetic field
strength in the shock region, hence producing radio synchrotron
emission \citep[e.g.][]{1998A&A...332..395E,2012arXiv1204.2455I}. Some
clusters show a relic and a radio halo, like the Coma cluster
\citep{1993ApJ...406..399G}, others, such as Abell 115 only host a
single relic \citep{2001A&A...376..803G}, a few objects have been
found where double relics are present \citep[e.g. Abell 1240 and Abell
  2345][]{2009A&A...494..429B}. Finally, in a couple of objects, both
double relics and a radio halo have been found \citep[e.g. CIZA
  J2242.8+5301][]{2010Sci...330..347V}. The combinations with which
halos and relics are found, and the number of these objects itself,
are likely to be affected by detection limits of current
instruments. In the last years, efforts have been made from both
observational and theoretical sides to better understand the
properties of radio relics. From a theoretical point of view, the
properties of merging shocks, and the acceleration mechanism required
to explain the radio emission, have been investigated in several works
\citep{pf06,2007MNRAS.375...77H,2009MNRAS.393.1073B,2009A&A...504...33V,va10kp,2011ApJ...735...96S,2011ApJ...734...18K,2012MNRAS.420.2006N,Vazza12,2012arXiv1204.2455I}.
Recently, \citet{2012MNRAS.420.2006N} have used hydrodynamic
simulations to predict the number of observable relics as a function
of the redshift. They find that either more relics with $z>0.3$ should
be found than known to date (only a couple have been observed so far),
or that the ratio $B/n_e$ of the magnetic field $B$ over the thermal
gas density, $n_e$, changes with $z$. The ``lack'' of relics at
$z>0.3$ can be however due to the lack of deep radio observations of
distant galaxy clusters. Investigating the properties of galaxy
clusters in this redshift range would hence provide information on the
physics of the ICM itself. \\ So far, the properties of radio halos
and relics have been mainly studied in galaxy clusters at low redshift
($z<0.3$). \citet{Giovannini09} have analyzed the radio emission from
a sample of galaxy clusters at $z<0.2$, finding that at the NVSS
detection limit, the percentage of galaxy clusters with diffuse
sources is in the range 6-9\% in clusters with X-ray luminosity
$L_x>10^{44}$ erg s$^{-1}$, and grows to $\sim$40\% when only clusters
with $L_x>10^{45}$ erg s$^{-1}$ are
considered. \citet{2008A&A...484..327V} have searched for the presence
of radio halos in a complete sample of X-ray luminous galaxy clusters,
finding that the fraction of clusters with a radio halo is $\sim$30\%
in the redshift range 0.2 - 0.4 (with most of the objects having $z
\leq 0.3$) increasing up to $\sim$40\% when clusters with $L_x>8 \cdot
10^{44}$ erg s$^{-1}$ are considered.\\ The aim of the present work is
to extend the sample of diffuse radio sources to $z > 0.3$, the epoch
when massive clusters formed.  X-ray selected samples compiled from
ROSAT data, such as REFLEX \citep{2004A&A...425..367B} and eBCS
\citep{2000MNRAS.318..333E} have probed the radio properties of galaxy
clusters in the local Universe
\citep[e.g.][]{2008A&A...484..327V,Giovannini09}. At higher redshifts
the MACS project has compiled the first large X-ray selected sample of
clusters that are both X-ray luminous and distant
\citep{Ebeling01}. It consists of 128 clusters at $z>0.3$. The MACS
catalogs of \citet{Ebeling07,Ebeling10} comprise all clusters
at $z>0.3$ with a nominal X-ray flux larger than $2 \, \cdot 10^{-12}$
erg s$^{-1}$ cm$^{-2}$ (0.1--2.4 keV) in the ROSAT Bright sources
catalog \citep{Voges99}, and 12 clusters at $z>0.5$ down to an X-ray
flux limit of $1\times 10^{-12}$ erg s$^{-1}$ cm$^{-2}$. A third MACS
subsample was recently published by \citet{2012MNRAS.420.2120M}.

We have inspected the archival Chandra/XMM-Newton data of the clusters
contained in the MACS catalogs, and we have selected those that show a
disturbed X-ray morphology. To start with, we have observed four
promising candidates that show hints of diffuse emission from the NVSS
\citep{NVSS}. However, there are further candidates for diffuse emission in the MACS sample that will be followed up in future work. \\ The paper is organized as follows: The observations and
the data reduction are described in Sec. \ref{sec:Obs}, followed by the results in Sec. \ref{sec:results}. We
discuss our results in Sec. \ref{sec:discussion}, and compare our observations with
mock radio images obtained with cosmological simulations. In
Sec. \ref{sec:correlations}, we present new correlations for the radio
halos and double-relic systems. Finally, we conclude in
Sec. \ref{sec:conclusions}.\\ Throughout this paper we assume a
concordance cosmological model $\rm{\Lambda CDM}$, with $H_0=$ 72 km
s$^{-1}$ Mpc$^{-1}$, $\Omega_M=$ 0.27, and $\Omega_{\Lambda}=$ 0.73.

\begin{table*} 
\caption{Observation details} %
\label{tab:radioobs}      
\centering          
\begin{tabular}{|c c c c c|}     
\hline\hline       
Cluster  & RA       &  DEC   &  $z$ &  Observation Date    \\
         & (J2000)  & (J2000)&    &                     \\
\hline       
MACSJ1149.5+2223  & 11 49 34.3 & 22 23 42  & 0.544$^{c}$ & 15-AUG-201 \\
MACSJ1752.0+4440  & 17 52 01.5 & 44 40 46  & 0.366$^{b}$ & 27-JUN-2011 \\
MACSJ0553.4$-$3342 & 05 53 26.7 & -33 42 37 & 0.431$^{d}$ & 15-AUG-2010 \\
MACSJ1731.6+2252  & 17 31 40.1 & 22 52 39  & 0.389$^{a}$ & 26-JUN-2011  \\
\hline
\hline
\multicolumn{5}{l}{\scriptsize Col. 1: cluster name; Col. 2 and 3:
  Right ascension and Declination (J2000) from
  RASS.}\\ 
\multicolumn{5}{l}{\scriptsize Col. 4: redshift ($a:$
  \citet{Ebeling10}, $b:$ \citet{2003MNRAS.339..913E},
  $c:$\citet{Ebeling07}, $d:$  \citet{2012MNRAS.420.2120M})}\\ 
\multicolumn{5}{l}{\scriptsize  Col. 5: Observation date.}
\end{tabular}
\end{table*}

\section{Radio observations}
\label{sec:Obs} 
\subsection{GMRT observations and data reduction}
The observations were carried out at the Giant Metrewave Radio
Telescope (GMRT) during June - August 2010.  We chose to perform the
observations at 323 MHz, as this frequency provides the best
compromise between sensitivity, resolution, and good sampling of the
short baselines. High resolution is crucial, since the emission from
discrete sources needs to be separated from the halo/relic emission. A
good coverage of the short spacings is also required to image the
diffuse emission. One Mpc corresponds to $\sim$3$'$ and 4$'$ at
redshift 0.5 and 0.3 respectively, much smaller than the largest
visible structure of the GMRT at 323 MHz (32$'$). This ensures that
radio emission with typical sizes of radio halos and relics can be
imaged. Each cluster was observed for 6h, recording both RR and LL
polarizations with 32 MHz bandwidth in 256 channels each. However, due
to high wind speeds a considerable fraction of the observing time was
lost in the June observations.  We used 8 sec integration time for a
more accurate removal of the Radio Frequency Interferences (RFI).  To
set the absolute flux density scale, the sources 3C286 and 3C48 (for
MACSJ1752.0+4440), 3C48 (for MACSJ0553.4-3342), 3C147 and 3C286 (for
MACSJ1149.5+2223), and 3C286 (for MACSJ1731.6+2252) were observed at
the beginning/end of the observing run for 10-15 min. The flux density scale
was set according to the \citet{PerleyTaylor} extension of the Baars
scale \citep{Baars77}. Flux density calibrators were also used to
correct the bandpass. To get a first calibration of the phases, we
alternated in each observing run between the target source (40 min
scan) and a phase-calibrator (5 min scan). As phase calibrators, we
observed 1635+381, 0447-220, 1120+143, and 1609+266 for the clusters ,
MACSJ1752.0+4440, MACSJ0553.4-3342, MACSJ1149.5+2223, and
MACSJ1731.6+2252 respectively.\\ The data were reduced using the NRAO
Astronomical Image Processing Systems (AIPS) package.  Data were
visually inspected to identify and remove strong and low-level RFIs.
Given the long initial scan on the flux density calibrator, and the
relatively low observing frequency, we selected a few channels (3 to
5) and solved every 8 sec for the complex gains on the flux density
calibrator sources before solving for the band response. This is done
because the amplitude and phases can change over 10-15 min at this low
frequency.  Hence, such effects have to be taken out before the
bandpass is computed over the whole calibrator scan (see also
\citealt{2011A&A...527A.114V}).  After this initial calibration, we
selected 10-15 channels free and/or cleaned from RFIs (usually
160-170) to normalize the bandpass. When 2 scans on the flux density
calibrators were present, at the beginning and at the end of the
observation, the bandpass solutions were computed separately for each
scan, to check that the band response was stable. A time interpolation
of the two was then applied to the data. After the bandpass
calibration, the channels at the edges of the band were removed
(usually the first and last 8 channels), and the remaining ones were
averaged down to 48 channels. We did not average further in order to
minimize the effects of bandwidth smearing.  We then performed a more
robust calibration of the flux density calibrator sources, this time
including all the 48 channels, and solving for the whole scan, to get
higher signal-to-noise solutions. Gains for the phase calibrators were
then computed, and their flux densities were bootstrapped from the
flux density calibrators. Solutions were interpolated and applied to the target
sources. Several cycles of phase self-calibration have been done on
the target source, followed by a final amplitude and phase self
calibration run to refine the antenna gains.\\ In order to account for
the effects of non-coplanar baselines, data were imaged using
``3-dimensional" techniques: 40-45 facets were used in the cleaning
procedure to cover an area equal to $\sim$ 2 times the primary beam,
with each facet being tangent to the celestial sphere at its centre.
This ensures that side-lobes from distant sources are subtracted from
the central field.\\ Different weighting schemes were applied to the
data to obtain images at different resolution. Such weighting schemes
are described in the section for individual clusters
(Sec. \ref{sec:results}).  Finally, images were corrected for the
primary beam attenuation.  The uncertainty in the calibration of the
absolute flux density scale is estimated to be $\sim$8\%. The errors reported
on the flux density measurements are computed as
\begin{equation}
dS=\sqrt{(rms \times \sqrt{N\_{beam}})^2 + (0.08 \times S)^2}
\label{eq:noise}
\end{equation}
where rms is the rms noise of the image, $N\_{beam}$ is the number of
independent beams in the region where the flux density, $S$, is measured.
\subsection{VLA observations and data reduction}
L-band observations of the cluster MACSJ1149.5+2223 have been retrieved
from the VLA data archive. The observations were performed in the C
array configuration. The source 3C286 was used as primary flux density
calibrator and as an absolute reference for the electric-field vector
polarization angle.  The source 0842+185 was observed as phase
calibrator, and together with the sources 0503+020 and 1125+261, they
were used as parallactic angle calibrators.  We performed standard
calibration and imaging using the NRAO Astronomical Imaging Processing
Systems (AIPS). Cycles of phase self-calibration were performed to
refine antenna phase solutions on target sources, followed by a final
amplitude and gain self-calibration cycle in order to remove minor
residual gain variations.  The uncertainty in the calibration of
the absolute flux density scale is estimated to be $\sim$ 5 \%.

\subsection{WSRT observations }
The cluster MACSJ1752.0+4440 was observed with Westerbork Synthesis
Radio Telescope at 13,18,21, and 25 cm, and the resulting images have
been presented in \citet{2012arXiv1206.2294V}. Here we use the 18 cm
observation, that is 24 hour long, and provides the best imaging of the
radio halo.  We refer to  \citet{2012arXiv1206.2294V} for details about the data calibration.  The polarization
was calibrated using 3C48 to remove instrumental polarization, and
3C286 as absolute reference for the electric vector polarization
angle.\\

Total intensity, I, and Stokes parameter Q and U images have been
obtained from VLA and WSRT images.  Polarization intensity
$P=\sqrt{U^2+Q^2}$, Polarization angle $\Psi=\frac{1}{2} \arctan (U,Q)$ and
fractional polarization $FPOL=P/I$ images were obtained from
the I, Q and U images. Polarization intensity images have been
corrected for a positive bias.

\begin{table*} 
\caption{X-ray properties }   
\label{tab:xray}      
\centering          
\begin{tabular}{|c c c c|}    
\hline\hline       
Cluster  & $L_X \; [r_{500}]$ & $L_X \; [r_{500}]$ & $kT$ \\
         &  [0.1 - 2.4 keV]  &  bolometric        &  keV\\
\hline       
MACSJ1752.0+4440  & 0.8$\cdot 10^{45}$$^{b}$  &    -               &   6.7$^{b}$     \\
MACSJ1149.5+2223  & 1.4$\cdot 10^{45}$$^{c}$  &     5.1$\cdot 10^{45}$$^{c}$   & 14.5$^f$ \\
MACSJ0553.4$-$3342  & 1.7$\cdot 10^{45}$$^{d}$   &     5.7$\cdot 10^{45}$$^{d}$    & 13.1$^e$  \\
MACSJ1731.6+2252  & 0.8 $\cdot 10^{45}$$^{a}$  &     2.5$\cdot 10^{45}$$^{a}$   & 7.5$^e$\\
\hline
\hline

\multicolumn{4}{l}{\scriptsize Col 1: cluster name; Col. 2: X ray
  Luminosity in the 0.1 - 2.4 keV Band; Col. 3: Bolometric X-ray
   }\\ \multicolumn{4}{l}{\scriptsize Luminosity; Col. 4: Cluster
  temperature $a:$ \citet{Ebeling10}, $b:$
  \citet{2003MNRAS.339..913E}, $c:$\citet{Ebeling07},
}\\ \multicolumn{4}{l}{ \scriptsize $d:$ \citet{2012MNRAS.420.2120M},
  $^e$:\citet{2008ApJ...682..821C}, $^f$:\citet{2000ApJS..129..435B} }
\end{tabular}
\end{table*}

\section{Results}
\label{sec:results}

\begin{figure*} 
\centering
\includegraphics[width=0.495\textwidth]{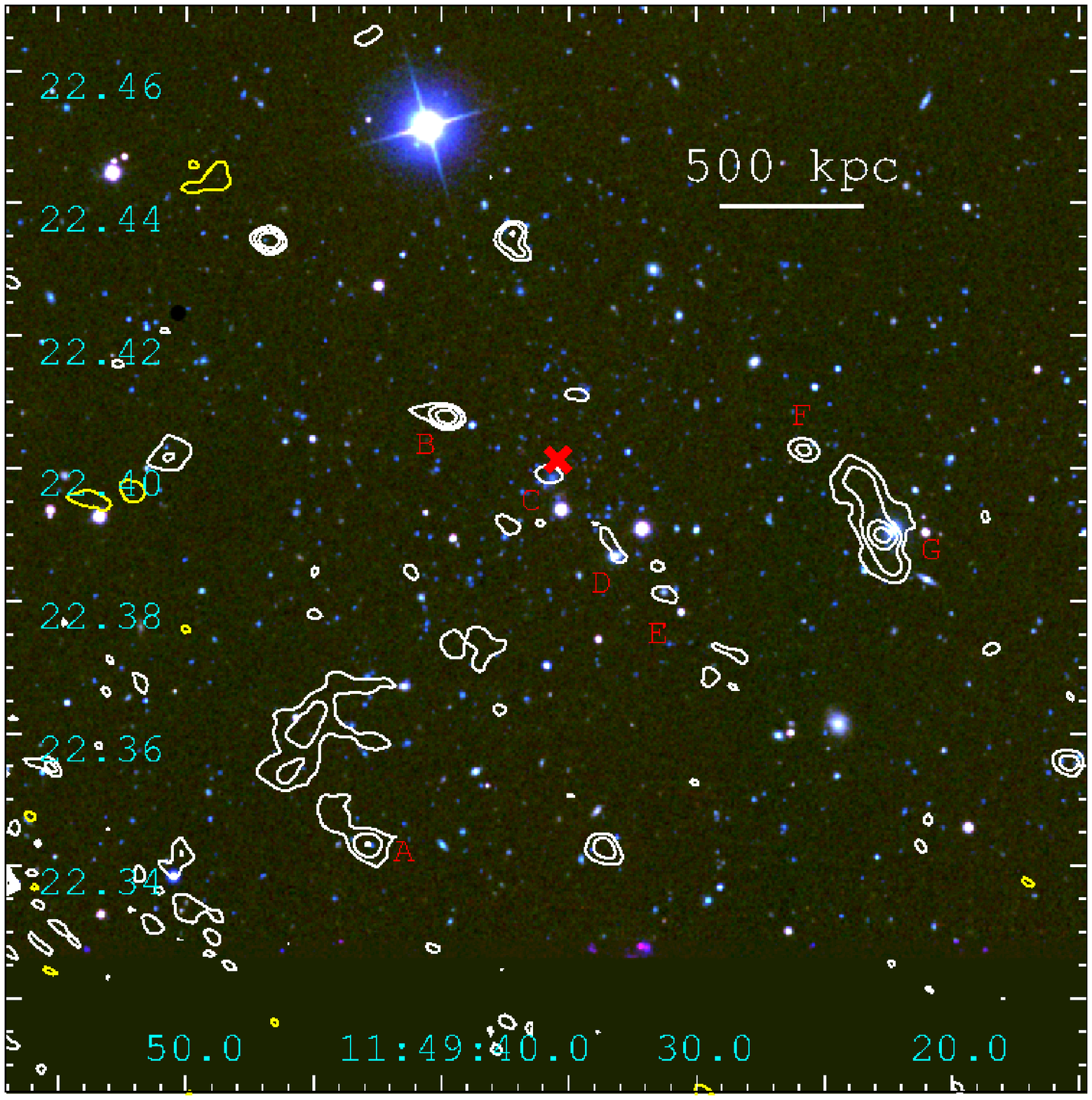}
\includegraphics[width=0.5\textwidth]{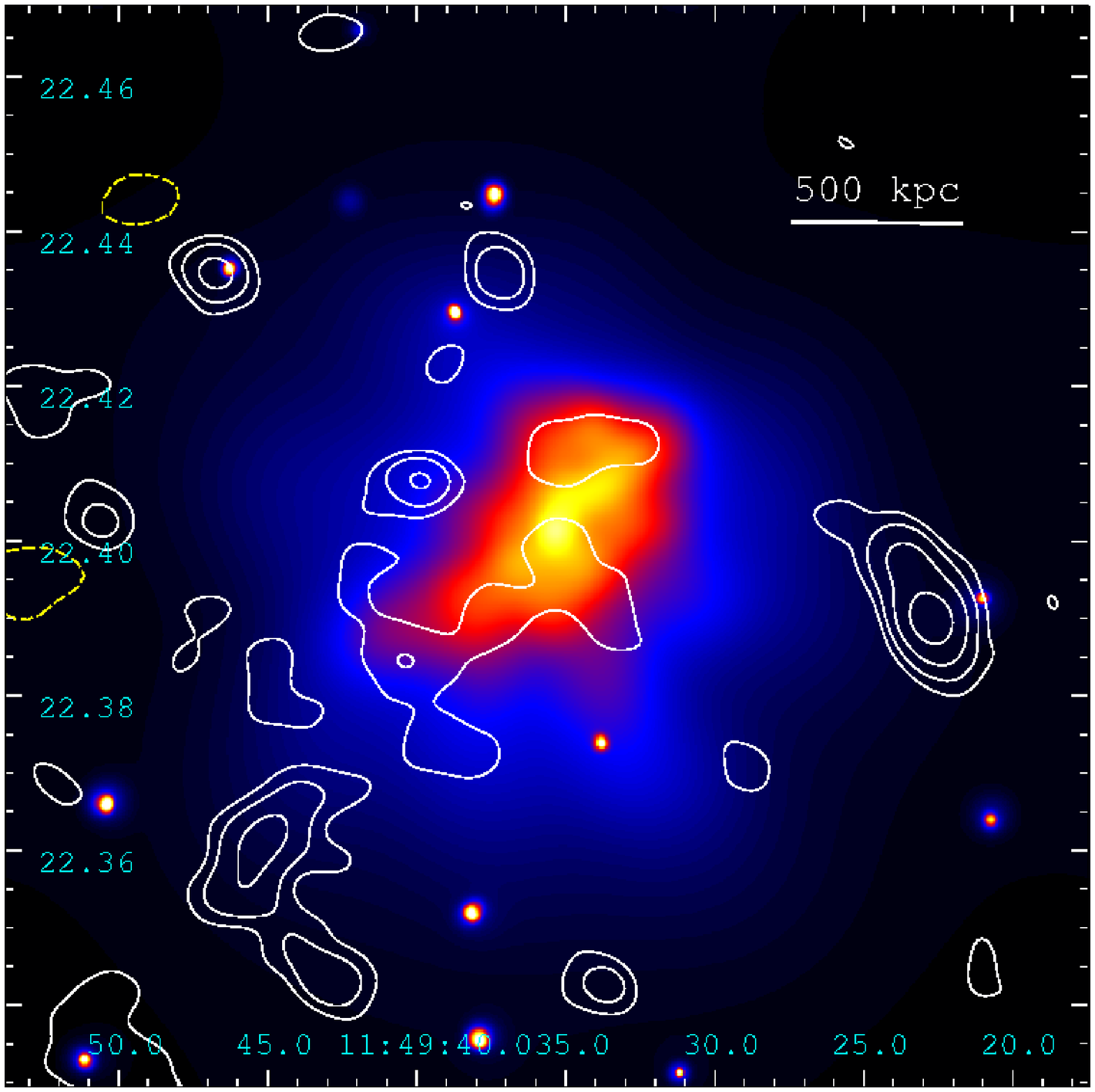}
\caption{{\bf MACSJ1149.5+2223} {\it Left:} the colour image shows the
  emission (bands r, g and i) from the SDSS Data release 7. Contours
  show the radio emission at 323 MHz from the GMRT, obtained with
  Robust$=$0 weighting. The restoring beam is 11.5$'' \,\times$
  7.8$''$. The rms noise is $\sigma \sim$ 0.2 mJy/beam. Contours start
  at 4$\sigma$ and are spaced by powers of 2. The -4$\sigma$ contour is
  displayed with dotted yellow lines. The red cross at the centre
  marks the position of the X-ray centre. Sources embedded in the
  relics' emission and within the cluster are marked by letters from A
  to F. {\it Right:} X-ray emission of the galaxy cluster as seen by
  {\it Chandra} ACIS-I detector, in the energy band 0.5-7 keV
  \citep{Ebeling07}. Contours show the radio emission at 323 MHz from
  the GMRT obtained with a Gaussian taper of the long baselines and
  Robust$=$1 weighting. The restoring beam is 24.5$'' \,
  \times$20.0$''$. The rms noise is $\sigma \sim$ 0.4
  mJy/beam. Contours start at 3$\sigma$ and are spaced by powers of
  2. The -3$\sigma$ contour is displayed with dotted yellow lines.}
\label{fig:macsj1149}
\end{figure*}

\begin{figure*}
\centering
\includegraphics[width=0.48\textwidth]{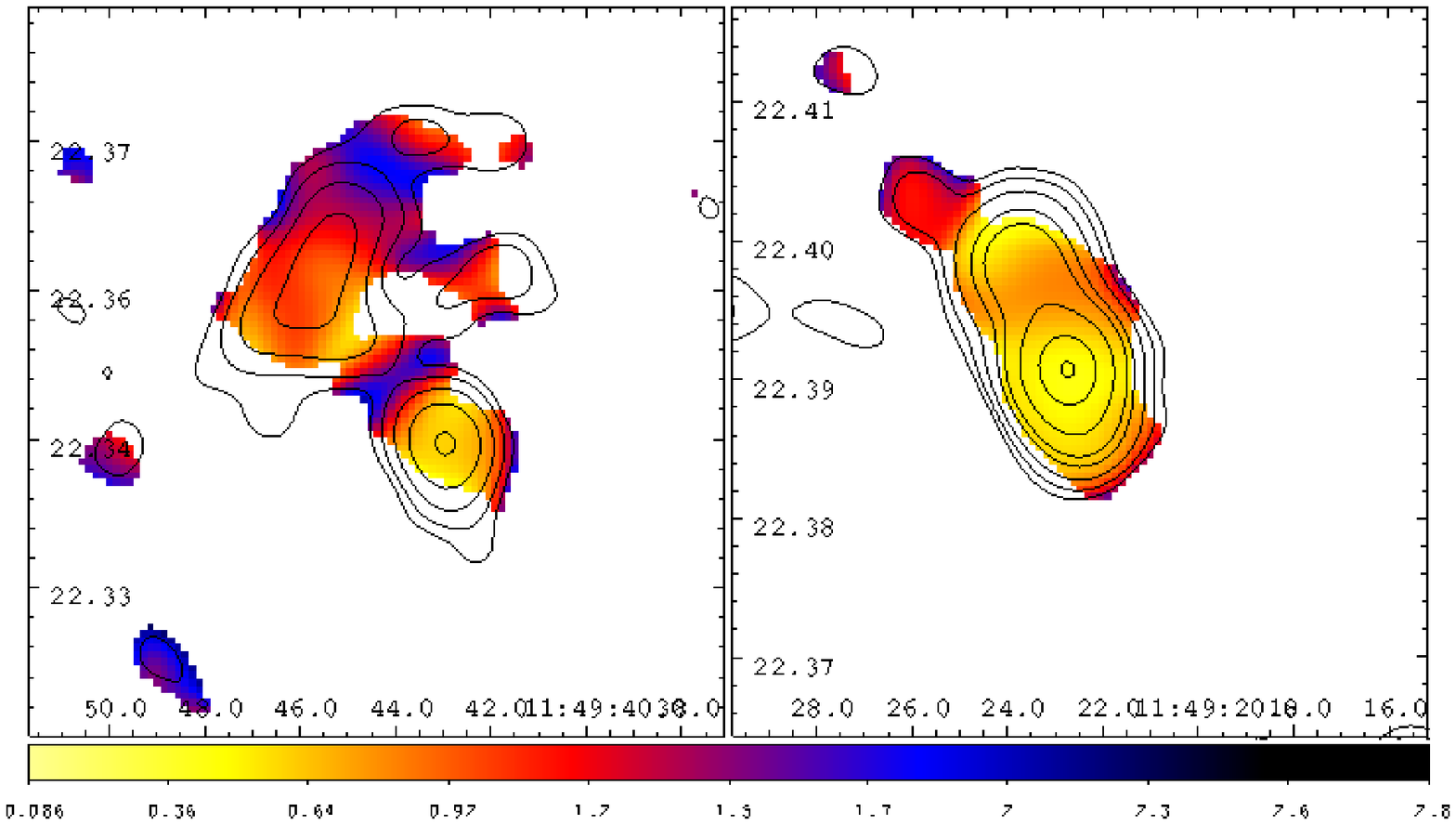}
\includegraphics[width=0.48\textwidth]{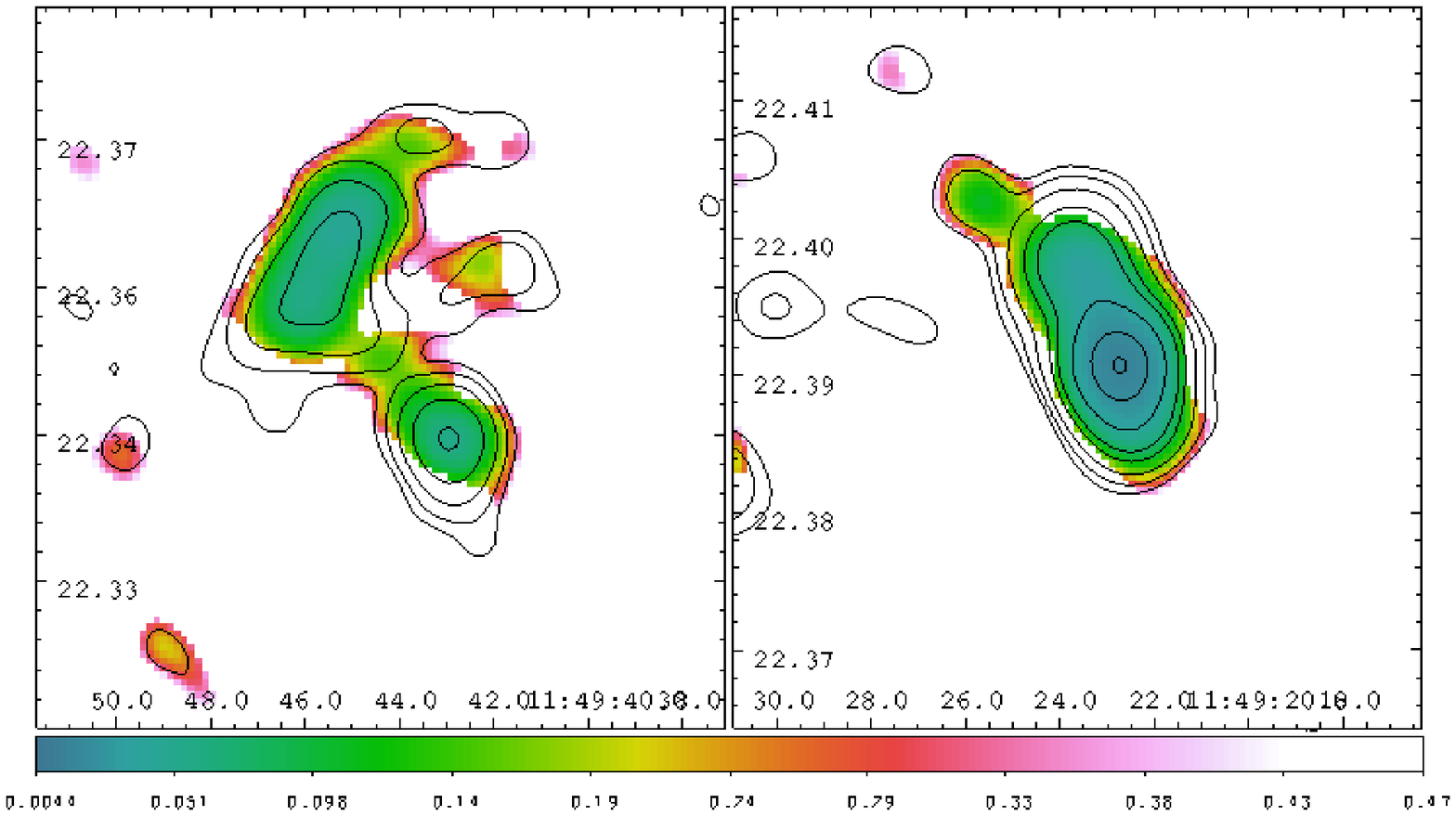}
\caption{{\bf MACSJ1149.5+2223} Spectral index images and errors for the
  relics in MACSJ1149.5+2223.  Contours start at 3$\sigma$ and are spaced by
  powers of 2.}
\label{fig:1149_spix}
\end{figure*}

\begin{figure*} 
\centering
\includegraphics[width=0.48\textwidth]{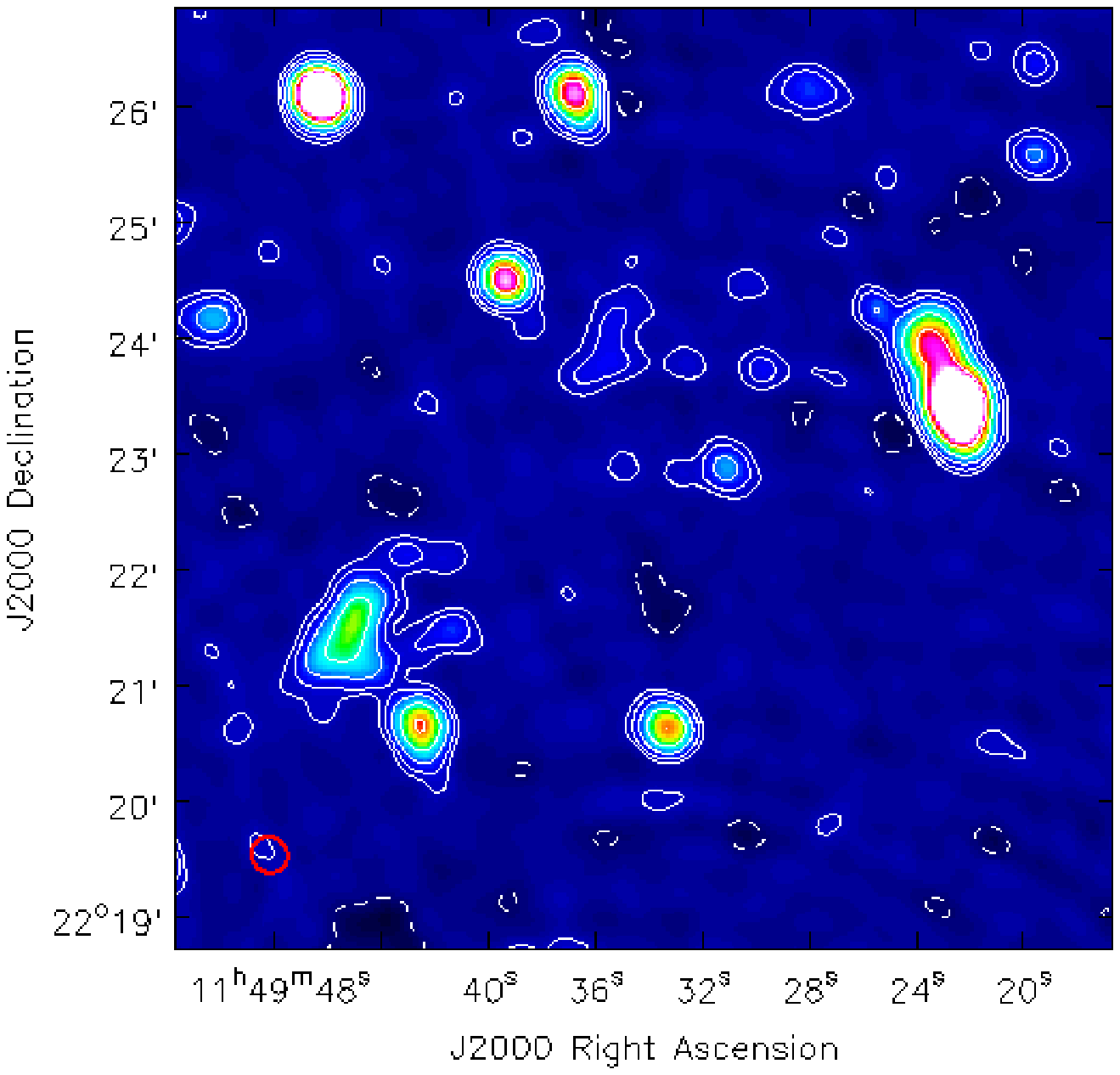}
\includegraphics[width=0.48\textwidth]{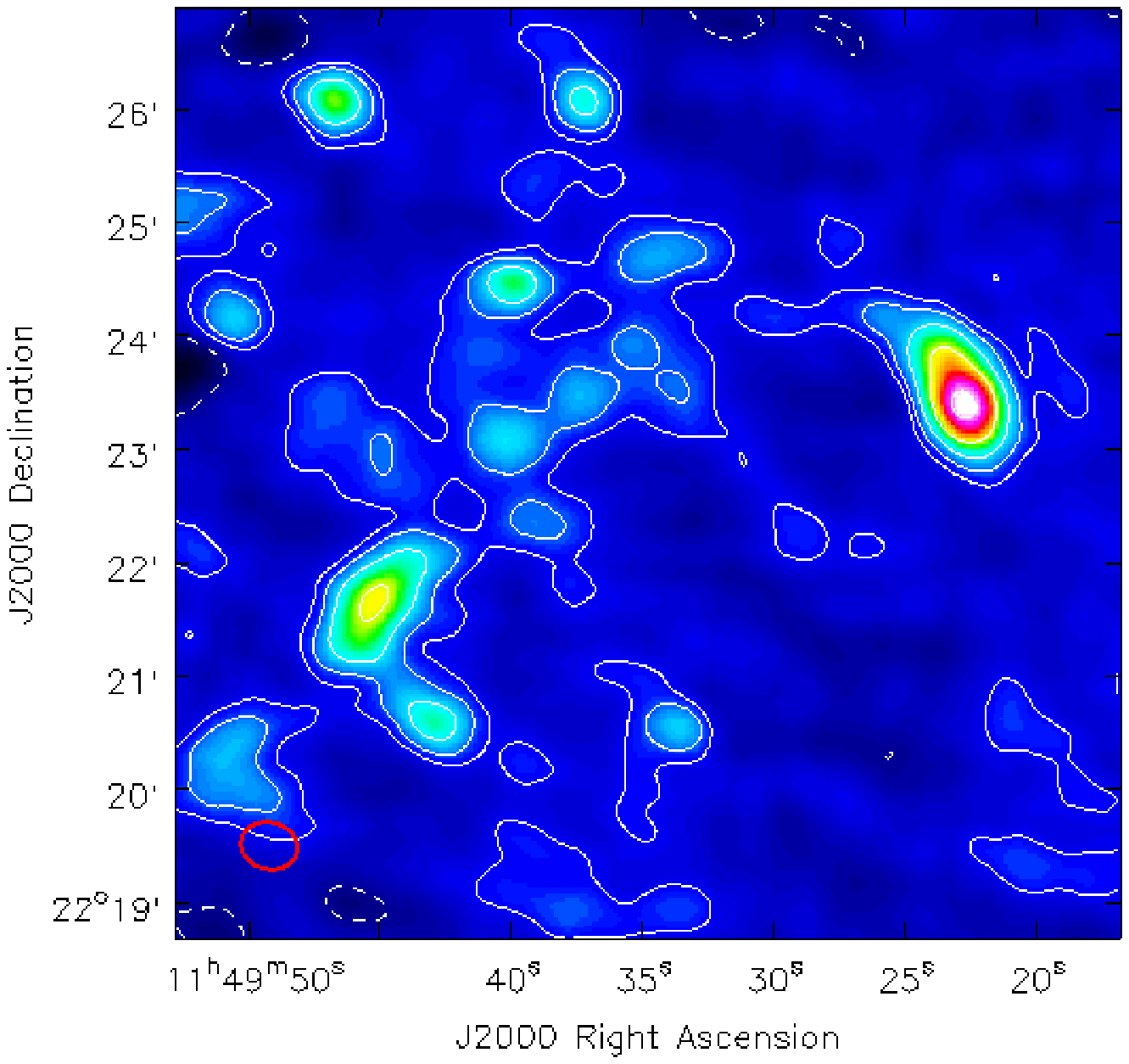}
\caption{{\bf MACSJ1149.5+2223} Radio emission from the cluster at 1.4
  GHz from VLA observations. The image is obtained with a Gaussian
  taper of the long baselines and Robust$=$1 weighting. The
  restoring beam is $\sim$20$''\times$20.0$''$, the rms noise level is
  30 $\mu$Jy/beam. Contours start at 3$\sigma$ and are spaced by
  powers of 2. The -3$\sigma$ contour is displayed with dashed
  lines. Right: Low-resolution image at 323 MHz obtained with natural
  weighting. The beam is $\sim$30$''\times$25$''$, The noise is 0.4
  mJy/beam. Contours start at 2 $\sigma$, the -2$\sigma$ contour is
  displayed with dashed lines.}
\label{fig:macs1149_2}
\end{figure*}

\begin{figure} 
\centering
\includegraphics[width=0.48\textwidth,angle=90]{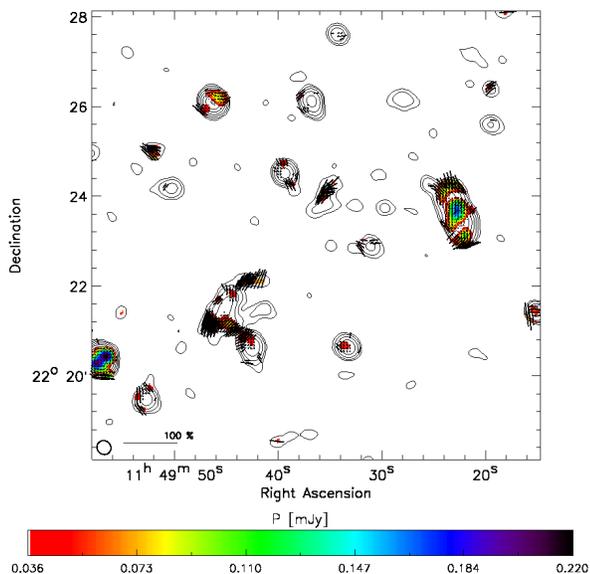}
\caption{Polarized emission in MACSJ1149.5+2223. Colours display the
  polarization intensity, while the lines display the direction of the
  electric field vector de-rotated by the Galactic Faraday depth. The
  length of the vectors is proportional to the fractional polarization
  For clarity, only one line every 3 pixels in plotted. Contours
  represent the VLA image obtained with natural weighting. The beam is
  23$''\times$16$''$, and is shown in red on the bottom left of the
  image. The noise is 15 $\mu$Jy/beam. Contours start at 3$\sigma$ and
  are spaced by powers of two. }
\label{fig:1149_pol}
\end{figure}

\subsection{MACSJ1149.5+2223}
\label{sec:1149}
The cluster MACSJ1149.5+2223 was observed with the Advanced Camera for Surveys on
board HST as part of a programme targeting all MACS clusters at
$z>0.5$ \citep{Ebeling07}.  A strong lensing analysis by
\citet{2009ApJ...707L.163S} identified seven multiple-image systems
and found the cluster to contain at least four large-scale mass
concentrations, making MACSJ1149.5+2223 one of the most complex
cluster lenses known to date. The contours of the luminous density,
mass density, and X-ray surface brightness, as derived from
Chandra observations \citep{Ebeling07}, show a net elongation toward
the SE-NW direction.  These observations have been used to place
constraints on the mass and structure of the cluster. The
fiducial model contains the main cluster halo plus three group-scale
halos.  \citet{Ebeling07} derived a total mass $M(\leq 500 \, \rm{kpc}) = (6.7 \pm
0.4) \times 10^{14} M_{\odot}$.\\ 

In the NVSS image, a radio source is located $\sim$130$''$ SE from the
cluster X-ray centre. This source is not detected in the FIRST survey
\citep{FIRST} because it is either resolved out (i.e. $\geq$ 2$'$), or
below the sensitivity limit. Another source is located at $\sim$
  3.7$'$ West from the cluster centre. It is detected in both FIRST
  and NVSS. We inspected the optical image of the cluster from the
  SDSS data release 7 (see Fig. \ref{fig:macsj1149}). There is an
  optical counterpart associated with the radio source, located at
  z$=$0.174 \citep{2002ApJS..143....1M}. The radio emission at 323
MHz is shown in Fig. \ref{fig:macsj1149}. Two radio sources are
located at $\sim$ 3$'$ and 3.7$'$ from the cluster X-ray centre, in SW
and SE directions, respectively.  Their extent is $\sim$2$'$ and
2.2$'$, which corresponds to 760 and 820 kpc at the cluster's
redshift.  We classify these sources are peripheral radio relics, or
radio ``gischt'', because of their location and morphology. We
  note that there is a foreground radio-galaxy projected onto the
  W-relic. Nonetheless, the morphology of its emission and the fact
  that no jets/hot-spots are visible in the higher-resolution image
  from FIRST indicate that the diffuse radio emission is likely not
  due to the radio galaxy but originates from the ICM. We also
  note that further emission from the W-relic is present with lower
  significance (see Fig. \ref{fig:macs1149_2}, right panel). Although
  deeper observations would be required to definitely rule out an AGN
  origin for this emission, we classify this source as a relic. The
two relics are clearly visible in the high-resolution image, obtained
with Robust $=$0 weighting (Fig. \ref{fig:macsj1149}, left panel).
The flux density of the discrete radio sources - labeled by letters in
Fig. \ref{fig:macsj1149} - are reported in Table \ref{tab:sources}. In
order to enhance the diffuse emission we have imaged the dataset with
a different weighting scheme (Robust$=$1 and uv-taper of the
long baselines).  The final image is shown in the right panel of
Fig. \ref{fig:macsj1149}. Further diffuse emission is visible between
the two relics which appears to be a radio halo as it follows
approximately the X-ray brightness of the cluster. The flux density of the
relics and of the candidate halo were computed from this image after
subtracting the flux density of the sources as measured from the higher
resolution image. To check the consistency of this approach, we have
selected the emission of the radio galaxies embedded in the cluster's
emission, and then subtracted the corresponding visibilities in the
Fourier plane, and re-imaged the dataset with a different weighting
scheme (Robust$=$1 and uv-taper of the long baselines). The two
approaches give consistent results. \\ The VLA 1.4 GHz images of
MACSJ1149.5+2223 are shown in Fig. \ref{fig:macs1149_2}. We imaged the
dataset with Robust$=$0 weighting, to isolate the emission from the
discrete radio galaxies, and with natural weighting to image the
diffuse emission. The two relics are detected at 1.4 GHz. However, due
to the poorer sampling of the inner uv-plane in comparison with GMRT
observations, their emission is not as well recovered as in the 323
MHz image. The candidate halo is barely visible in the 1.4 GHz
image. Details about the relics and the halo are reported in Table
\ref{tab:radio}.

\subsubsection{A spectral index study of the radio relics}

Two radio relics are detected in this cluster at both 323 MHz and 1.4
GHz. The relics are located at the edges of the X-ray emission. The
major axes of the relics are oriented perpendicular to the cluster's
X-ray elongation, that is likely tracing the main merger axis of this
complex system.  However, the relic positions appear to be misaligned with
the main merger axis, which is unique in double relics observed so
far. This peculiar feature is not reproduced by simple simulations
of binary mergers. We argue that it could be a sign of additional
mergers. \\ In order to derive the integrated spectral indices of the
relics' emission, we computed the radio brightness in the area where
relics are above the 3 $\sigma$ level in both VLA and GMRT
images. This cut is set by the VLA image, and thus the resulting
spectral indices will be representative of the flatter spectra
contributing to the relic emission.  The integrated spectral indices
are $\sim$0.75 $\pm$ 0.08 and $\sim$1.15 $\pm$ 0.08 for the Western
and Eastern relic, respectively.  After imposing a common uv-range cut
in the GMRT and VLA datasets, the two were reimaged using uniform
weighting and a tapering of the long baselines. Thus we take into accounr the different sampling of the uv-plane resulting from
observations performed with different arrays, at the cost of losing
part of the diffuse emission that is better sampled in the inner
uv-plane of the GMRT observations.  Both images were convolved to the
same restoring beam of 20$''$. The VLA and GMRT absolute position
differ by each other by few arcsec. The geometry itself is distorted,
so that there is not a single shift in RA or DEC, or a rotation of the
coordinates, that is able to compensate for that. This is most likely
caused by the self-calibration process in which the absolute position
of the sources is lost. This affects particularly the GMRT
observations since ionospheric effects can cause a shift in the
absolute positions of the sources. We have then shifted the VLA map to
fix the positions of sources A (for the Eastern relic) and F-G (for
the Western relic) separately. The shifts required to match the
position of the sources were 6$''$ in RA and 2$''$ in DEC for the
Eastern relic, and 4$''$ in both RA and DEC for the Western relic,
respectively. Only pixels above 2$\sigma$ in both images were
considered for the spectral index map. Furthermore, pixels in the
spectral index map with a signal-to-noise ratio smaller than 3 were
blanked in the final spectral index image.\\ The resulting
spectral-index maps are displayed in Fig. \ref{fig:1149_spix} together
with the spectral index error maps. In the Eastern relic, the spectral
index distribution shows flat values in the outer Eastern region
($\alpha \sim$0.6) and steepens toward the North-West, up to $\alpha
\sim$1.8.  The spectral index over the Western relic is more uniform,
with values going from $\alpha \sim$0.7 to $\sim$1.1, and no trend is
detected towards the cluster centre.\\ A steepening of the spectral
index is expected in the case of outgoing merger shock waves, and
observed in some relics
\citep[e.g.][]{2009A&A...494..429B,2010Sci...330..347V}. In
MACSJ1149.5+2223, we do not detect such a clear trend. However, in
this case the dynamic state of the ICM is more complex since X-ray and
lensing observations indicate the presence of several groups.
Due to the poor match of the absolute and relative positions for the
two datasets, any further discussion would be too speculative.

\subsubsection{Polarization properties}
 From U and Q Stokes maps, the polarized intensity and polarization
 angle maps were produced. We then derived the fractional polarization
 map by dividing the polarized intensity by the total intensity map.
 Pixels with noise larger than 10\% were blanked.  The relics show a
 mean polarization of $\sim$5\%. Due to the beam size, it is likely
 that beam-depolarization occurs, so that this value should be taken
 as a lower limit to the intrinsic polarization. We note that 20$''$
 corresponds to $\sim$100 kpc at the cluster redshift. The Galactic
 contribution to the observed polarization was estimated using the
 reconstructed map of the Galactic Faraday sky
 \citep{oppermann11}. The Galactic Faraday Depth at the location of
 the cluster is 0.68 $\pm$ 4.45 rad m$^{-2}$. The polarization vectors
 have been rotated accordingly. At such large distances from the
 cluster centre, the Rotation caused by the ICM is likely not
 significant, hence we assume that once the vectors are rotated by the
 Galactic FR, the result is the intrinsic polarization of the
 sources.  The polarization image is shown in
 Fig. \ref{fig:1149_pol}. The orientation of the vectors corresponds
 to the direction of the electric field vectors. In both relics, the
 magnetic field orientation is roughly aligned with the relics' main
 axis.  as found is some of the relics observed so far
 \citep[e.g.][]{2009A&A...494..429B,2010Sci...330..347V}, and in
 agreements with theoretical expectations.

\subsubsection{The radio halo}
The GMRT 323 MHz image shows low-brightness emission in the cluster
centre, covering the region in between the two relics. The brightest
part of this emission is detected at 3$\sigma$, although we find
indication of a larger and more extended emission with a lower
significance (2$\sigma$ above the noise level). To enhance the diffuse
emission, in Fig. \ref{fig:macs1149_2} we display a GMRT image
obtained with natural weighting, and overlay the first contour at
2$\sigma$ level. The shape of the radio emission roughly resembles the
X-ray brightness distribution of the cluster
(Fig. \ref{fig:macsj1149}). \\ The radio halo is barely visible in the
VLA 1.4 GHz image, probably because of the poorer sampling of the
shortest baselines. Nonetheless, positive residuals are visible in
this image as well. In order to give an estimate of the radio halo
power and spectral index, we have integrated the radio-flux density in the VLA
and GMRT images, in the area where the halo emission is detected at
the 2$\sigma$ level. The values are noted in Table
\ref{tab:radio}. Based on these values, the spectral index of the
  radio halo is $\alpha=2.1 \pm 0.3$. The spectral index value has to
  be considered a lower limit to the real spectral index since
  only the brightest knots are detected in the VLA map.
  The errors on the flux densities were computed using the formula in
  Eq. \ref{eq:noise} assuming 10\% errors on the flux density scale for both
  VLA and GMRT images, due to the lower significance of the
  detections. If we
only consider the area where the radio halo is detected at 3$\sigma$
in the GMRT image, we derive a similar value of $\alpha \sim $2.2$\pm$0.2. If
confirmed, this would be the halo with the steepest spectrum known so
far, and one of the very few with $\alpha >$1.6. Such steep spectrum
halos indicate synchrotron aging, difficult to explain with hadronic
models (\citealt{2008Natur.455..944B} and
\citealt{2009ApJ...699.1288D}). However, we note that deeper
observations at 323 MHz and a better uv-coverage at 1.4 GHz are
required to confirm the steep spectral index.\\

\begin{figure*}
\centering
\includegraphics[width=0.485\textwidth]{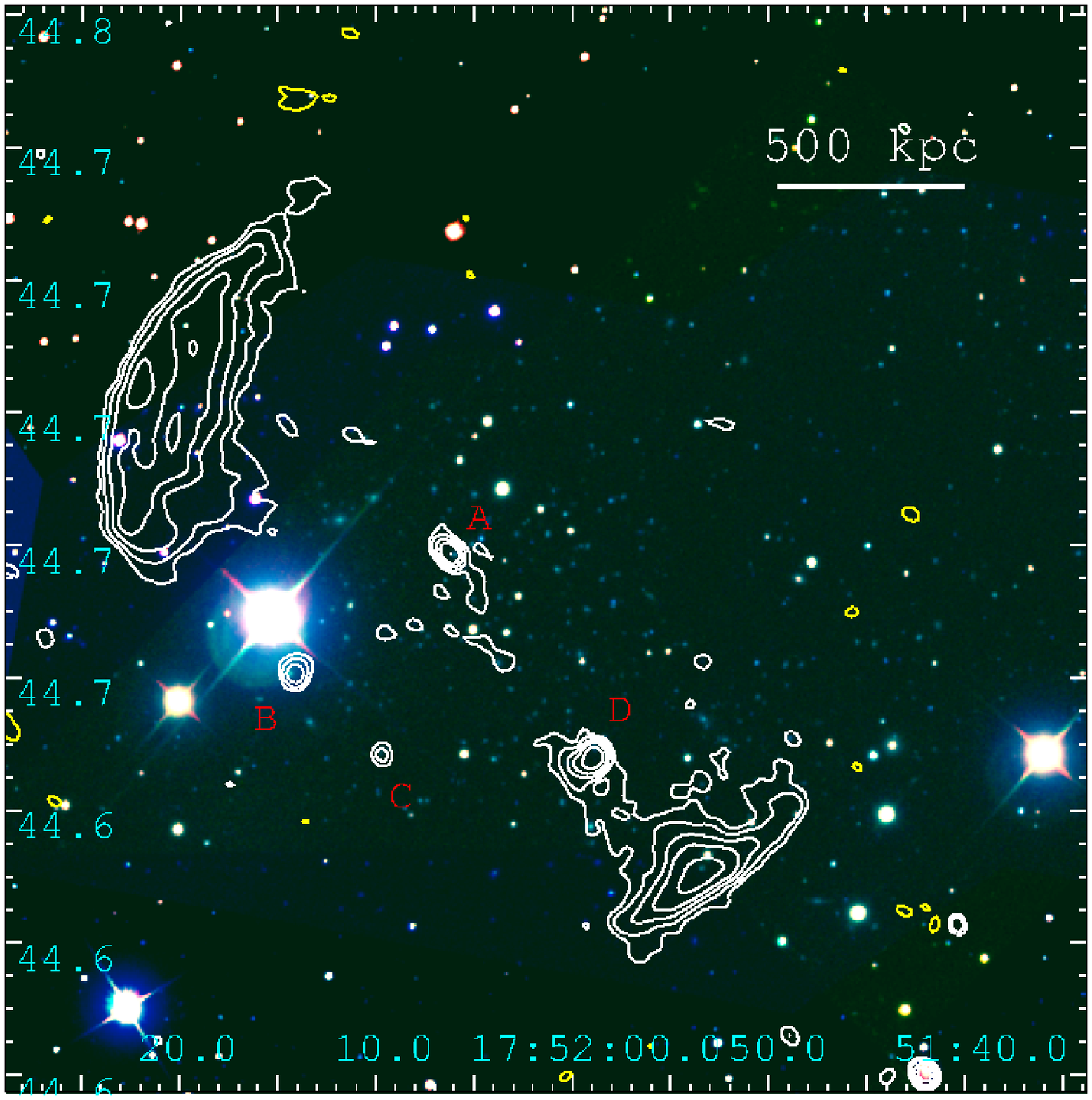}
\includegraphics[width=0.5\textwidth]{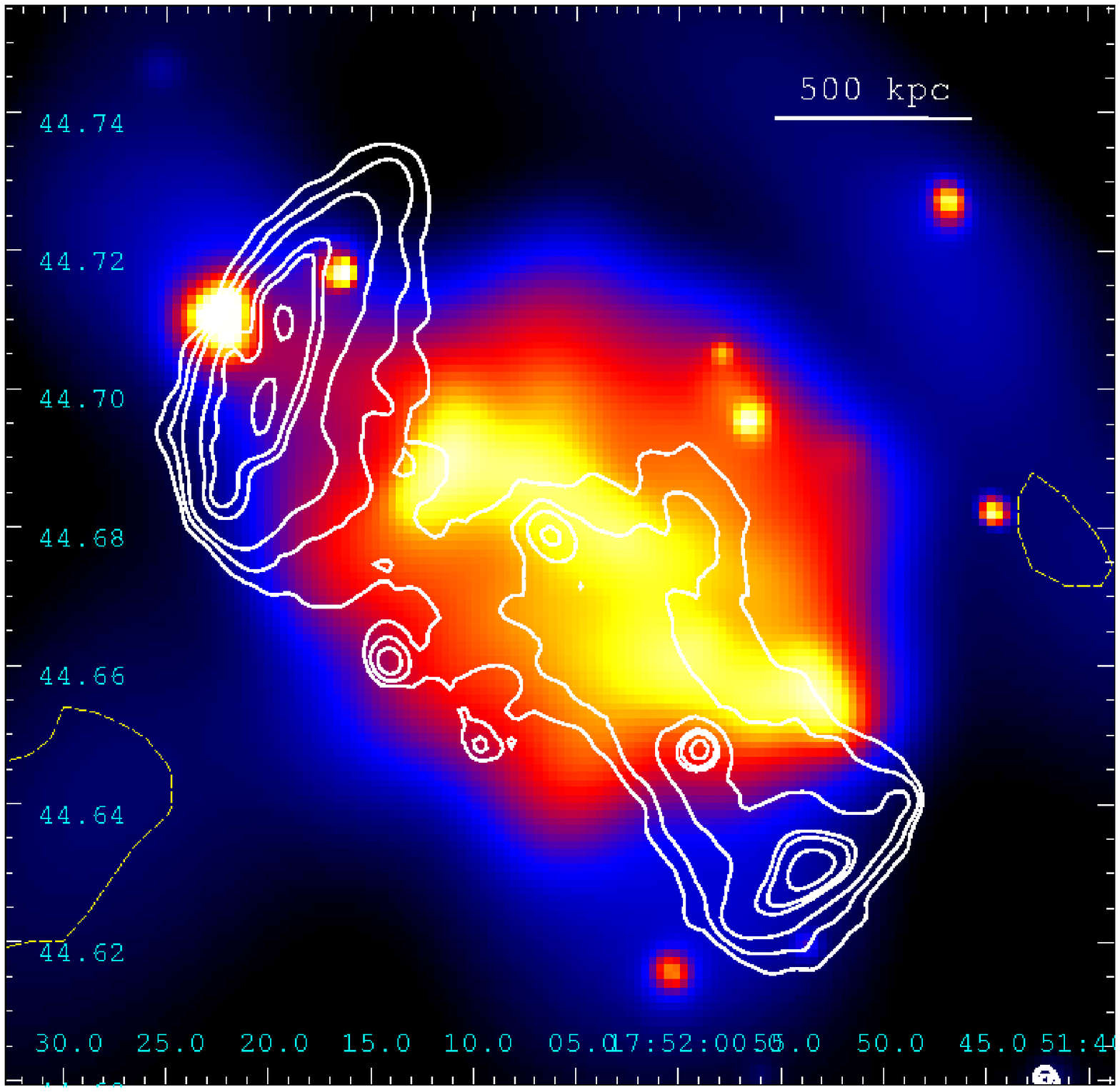}
\caption{{\bf MACSJ1752.0+4440} {\it Left:} the colour image shows the
  emission (i, r and g band) from the SDSS data release 7 . Contours
  show the radio emission at 323 MHz from the GMRT, obtained with
  Robust$=$0 weighting. The restoring beam is 9.9$'' \times$
  8.0$''$. The rms noise is $\sigma \sim$ 0.2 mJy/beam. Contours start
  at 3$\sigma$ and are spaced by powers of 2. The -3$\sigma$ contour is
  displayed with dotted yellow lines. The red crosses at the centre
  mark the position of the X-ray peaks.  {\it Right:} X-ray emission
  of the galaxy cluster as seen by {\it XMM-Newton} in the energy band
  0.2-12 keV. Contours show the radio emission at 323 MHz from the
  GMRT obtained with natural weighting. The restoring beam is $13.7'' \,
  \times$ 10.2$''$. The rms noise is $\sigma \sim$0.8 mJy/beam. Contours
  start at 3$\sigma$. Given the uniform brightness of the diffuse
  emissions, contours here are plotted at
  [0.0025,0.003,0,004,0.008,0.01,0.0015,0.003] mJy/beam.  The
  -3$\sigma$ contour is displayed with dotted yellow lines.}
\label{fig:macsj1752}
\end{figure*}

\begin{figure*}
\centering
\includegraphics[width=0.49\textwidth]{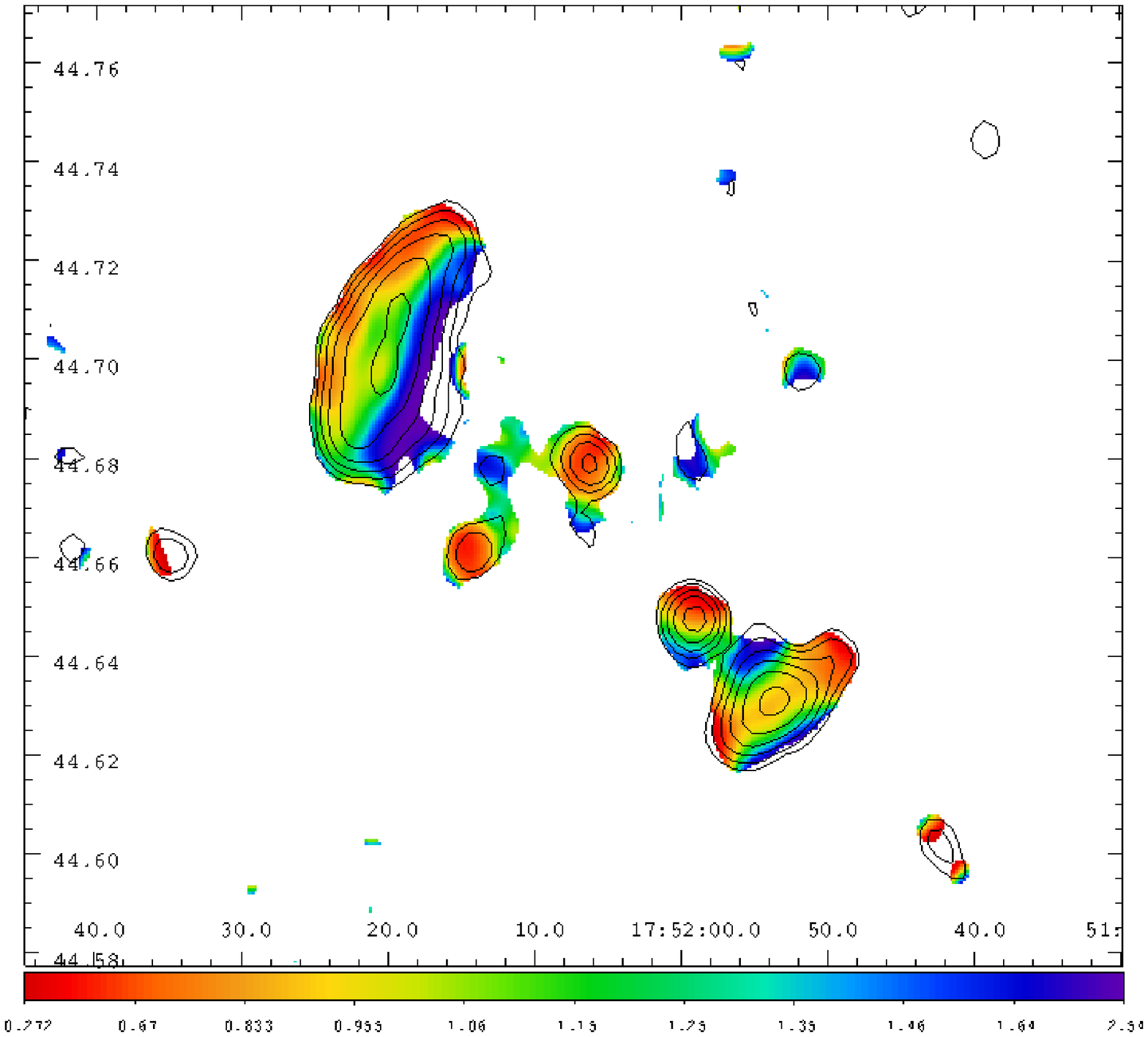}
\includegraphics[width=0.49\textwidth]{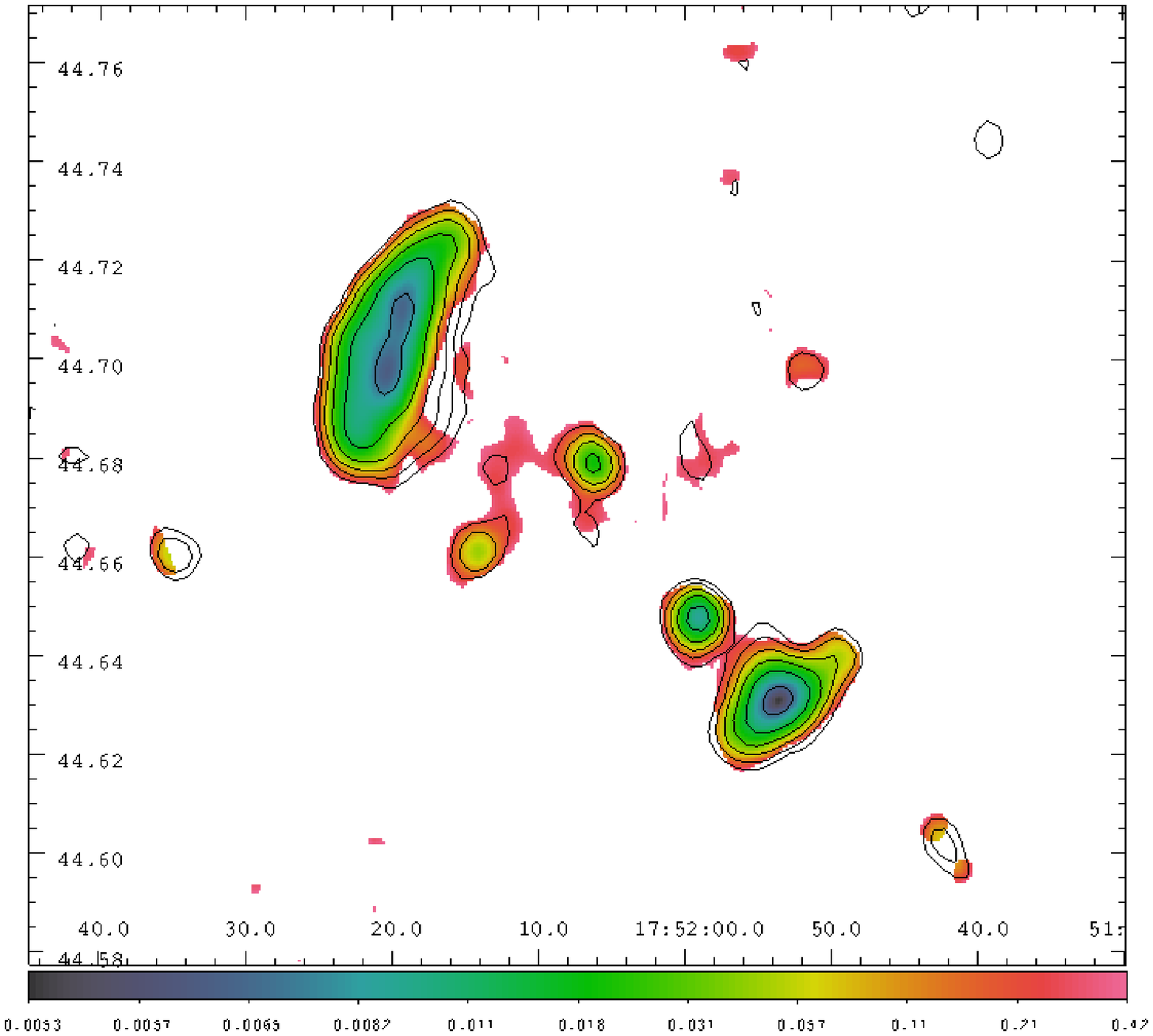}
\caption{{\bf MACSJ1752.0+4440} Colours: spectral index (left) and errors
  (right).  Contours are drawn from the GMRT image. Details about the
  imaging parameters are reported in Sec. \ref{sec:1752_spix}. The
  beam is $\sim$24$''\times$10$''$. The noise is 0.3
  mJy/beam. Contours start at 3$\sigma$ and are spaced by powers of 2. }
\label{fig:1752_spix}
\end{figure*}

\begin{figure}
\centering
\includegraphics[angle=-270,width=0.5\textwidth]{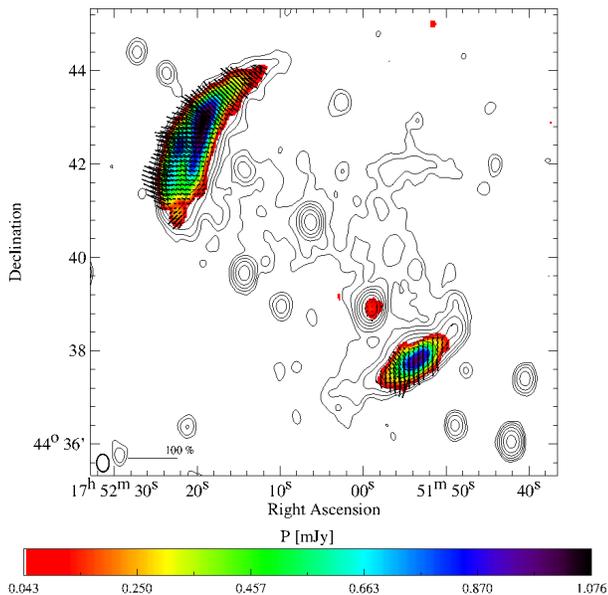}
\caption{{\bf MACSJ1752.0+4440} polarized emission. Colours display
  the polarization intensity, while lines display the direction of the
  E vector de-rotated by the Galactic Faraday depth. The lines are
  proportional to the fractional polarization. For display purposes,
  only one vector every 3 pixels is plotted. Contours are drawn from
  the WSRT image, obtained with natural weighting. The beam size
  is plotted in the bottom-left corner of the image. Contours start at
  3$\sigma$ and are spaced by powers of 2.}
\label{fig:1752_pol}
\end{figure}

\subsection{MACSJ1752.0+4440}
This is the only cluster among those presented here, that has not been
observed with Chandra. The XMM\_Newton image shows the cluster to be
elongated in the SE-NW direction, indicative of an on-going merger in
that direction.  \citet{2003MNRAS.339..913E} point out that it is a
very promising candidate to host two radio relics, as seen from NVSS
and WENSS radio images. Recently,  \citet{2012arXiv1206.2294V}
have confirmed the presence of such diffuse emission with Westerbork
observations.\\ Our GMRT observation confirms the presence of the two
relics and of a low-brightness radio halo (Fig. \ref{fig:macsj1752}).
The relics have a similar extension and morphology at 323 MHz and 1.7
GHz, although 1.7 GHz observations show a larger extent for the
SW relic. We consider the centre of the cluster to be the midpoint
between the two brightest X-ray cores.  The distance of the NE and SW
relics from the cluster centre is approximately 1130 kpc and 910
kpc. Their main properties are listed in Table \ref{tab:radio}.  A
very extended radio halo is detected in-between the two relics.  The
emission of the radio halo at 323 MHz is more extended than at 1.7 GHz
 \citep{2012arXiv1206.2294V}. The total extent in the NW-SE
direction seems wider at 323 MHZ, although a region located in the NE
part is only detected at 1.7 MHz. This indicates that particles with
different spectra are involved in the radio halo emission and/or that
the magnetic field strength is non-uniform.  The radio emission
follows the elongation of the X-ray emission with a rather flat
brightness distribution. The halo is elongated along the NE-SW
direction, with a fairly irregular morphology.  It is worth noting
that there is no clear correlation between radio and X-ray surface
brightness in this cluster: the brightest parts of the X-ray emission
do not correspond to the brightest radio emission. If we are observing
the first major-merger encountered by the galaxy cluster through its
life, this feature could be attributed to the less regular and
symmetric atmosphere that characterizes the ICM at this stage.  Details
about the radio halo properties are listed in Table \ref{tab:radio}.

\subsubsection{Spectral index analysis}
\label{sec:1752_spix}
Using WSRT data at 18 cm and the GMRT data, we derive the integrated
spectral indices of the radio relics and of the halo.  In order to
compute the spectral indices, the flux densities have been computed in the
area where both WSRT and GMRT images show emission above 3$\sigma$.
For the relics this is not critical since their sizes are very similar
at the two frequencies. However, the radio halo appears on average
more extended at 323 MHz, but some regions in the North are visible
only in the WSRT radio image.  The spectral indices we derive are 1.21
$\pm$ 0.06 and 1.12 $\pm$ 0.07 for the NE and SW relic,
respectively. The radio halo has a spectral index of 1.33 $\pm$
0.07.\\ WSRT 1.7 GHz and GMRT 323 MHz observations have been used to
produce a spectral index image of the radio emission. The two datasets
were re-imaged using the same UV-range, uniform weighting and a
Gaussian uv-taper of the long baselines. This ensures that different
samplings of the uv-plane, due to the different instruments, are
properly taken into account. The downside is that the extended weak
diffuse emission is partly lost. In particular, the radio halo has a
uniform brightness at 323 MHz and a large angular size of $\sim$5.5$'$
so that only the brightest knots are visible in the GMRT map.  The
spectral index image is shown in Fig. \ref{fig:1752_spix}. Only pixels
above 2$\sigma$ in the individual images were considered for the
spectral image, and furthermore, pixels with $S/N$ ratio smaller than
3 were blanked out in the output image. The NW relic shows a clear
steepening of the spectral index going from the cluster periphery
toward the centre. Such a clear trend has been detected before only in
the northern relic of CIZAJ2242.8+5301
\citep{2010Sci...330..347V}. The spectral index values range from
$\alpha \sim$ 0.6 to $\alpha \sim$ 2.5. The SW relic shows also a net
gradient along the relics' main axis. The flattest part of the
spectral index is not located at the edge of the relic emission, but
more central, and coincide with the peak of the radio emission. This
could suggest that the possible shock responsible for the relic origin
is propagating oriented at a tiny angle toward the observer. If that
is the case, the freshly accelerated particles are seen in the
foreground of the already aged ones. The spectral index values range
from $\alpha \sim$0.8 to $\alpha \sim$1.7 towards the edges of the
relic.
\subsubsection{Polarization}
WSRT observations provide polarization information for the cluster
radio emission.  From Stokes U and Q images we have derived the
polarization intensity image, the image of the polarization angle and
the fractional polarization map as explained in Sec. \ref{sec:1149}.
The Galactic Faraday rotation was corrected as explained in the same
Section. MACSJ1752.0+4440 is at Galactic longitude 71$^{\circ}$ and
latitude 29$^{\circ}$. The galactic Faraday depth is estimated to be
37.65$\pm$11.15 rad m$^{-2}$ \citep{oppermann11} and the polarization
angles have to be rotated accordingly.  The polarization image is
displayed in Fig. \ref{fig:1752_pol}. It shows that the NW relic is polarized on
average at a 20\% level, with values going up to 40\%. The polarization
is higher in the relics' external region, and decreases gradually
towards the cluster centre.  The SW relic is polarized at a 10\% level, on average, with
values up to 40\% in the external region. The high level of
polarization suggests an ordered magnetic field at the relic
position. It is possible that turbulent motions
developed after the shock passage contribute to randomize the magnetic
field, hence reducing the observed polarized flux.

\begin{figure*}
\centering
\includegraphics[width=0.51\textwidth]{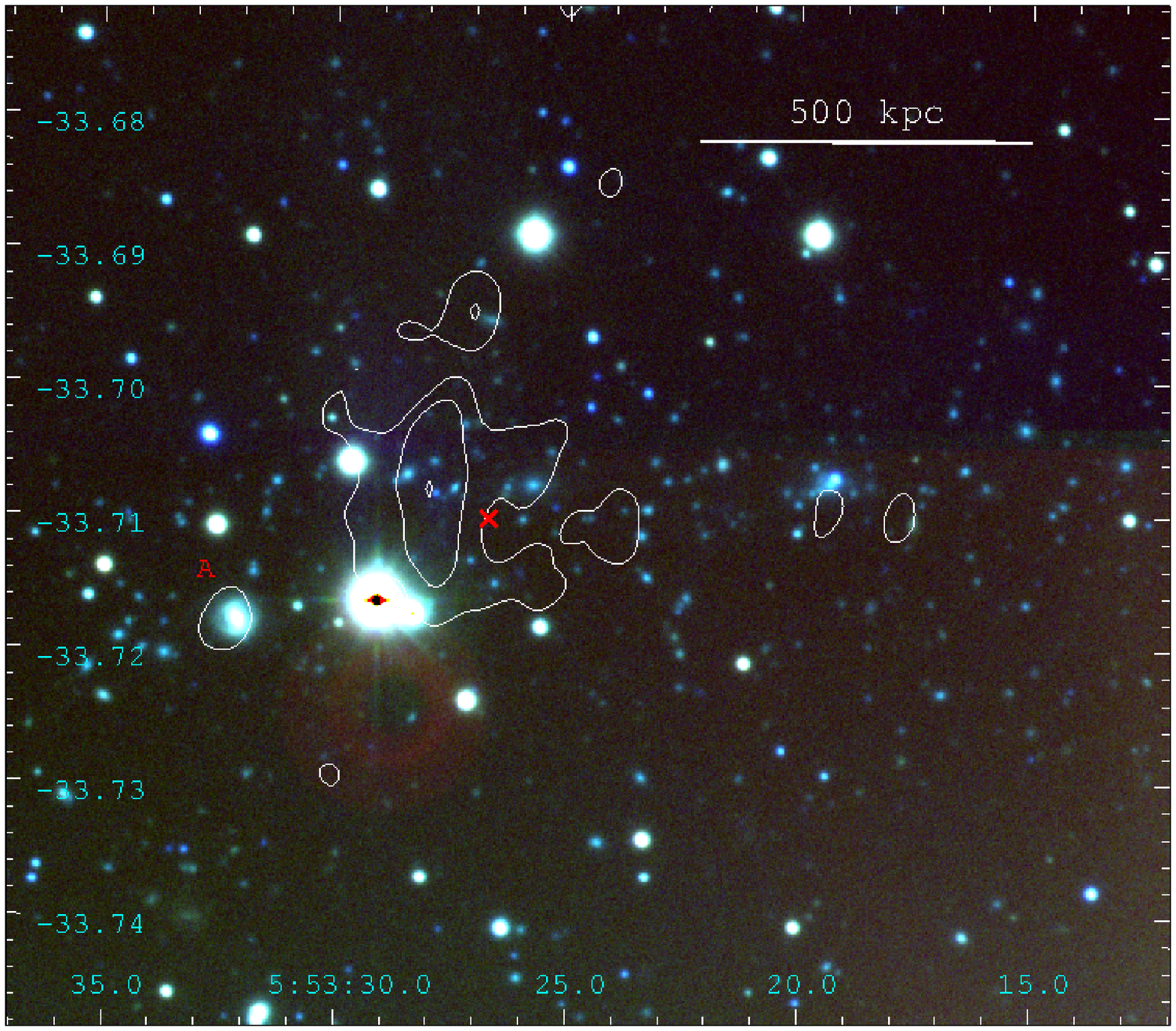}
\includegraphics[width=0.45\textwidth]{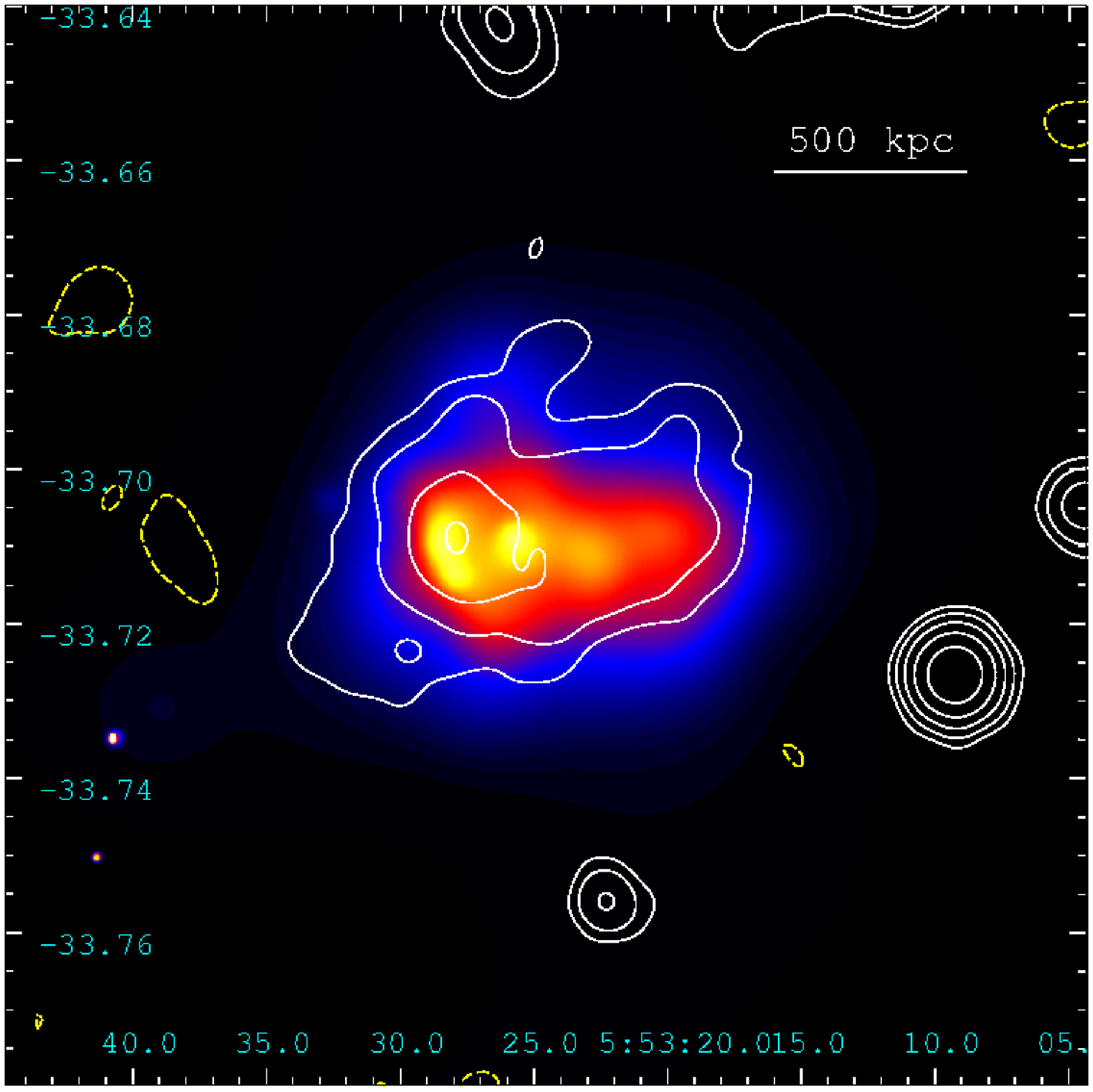}
\caption{{\bf MACSJ0553.4-3342} {\it Left:} the colour image shows the
  emission (v, r, and i bands) from the UH2.2m telescope
  \citep{2012MNRAS.420.2120M}. Contours show the radio emission at 323
  MHz from the GMRT, obtained with Robust$=$0 weighting. The restoring
  beam is 13.7$'' \,\times$ 8.6$''$. The rms noise is $\sigma \sim$
  0.14 mJy/beam. Contours start at 3$\sigma$ and are spaced by powers
  of 2. The -3$\sigma$ contour is marked by a yellow line. The red
  cross at the centre marks the position of the X-ray centre. The
  source embedded in the halo emission is marked by letter A. {\it
    Right:} X-ray emission of the galaxy cluster as seen by {\it
    Chandra} ACIS-I detector, in the energy band 0.5-7 keV
  \citep{Ebeling07}. Contours show the radio emission at 323 MHz from
  the GMRT obtained with a Gaussian taper of the long baselines and
  Robust$=$3 weighting. The restoring beam is 23.2$'' \,
  \times$21.0$''$. The rms noise is $\sigma \sim$ 0.25
  mJy/beam. Contours start at 3$\sigma$ and are spaced by powers of
  2. The -3$\sigma$ contour is marked by a yellow line.}
\label{fig:macsj0553}
\end{figure*}
\subsection{ MACSJ0553.4-3342}
 MACSJ0553.4-3342 is a disturbed cluster, as evident from both optical
 and X-ray images \citep{2012MNRAS.420.2120M,Ebeling10}. According to
 \citet{2012MNRAS.420.2120M} it is undergoing a binary head-on merger,
 and it is currently observed after the core passage. Two X-ray peaks,
 clearly visible in the image, are likely to be the cores of the two
 clusters. The X-ray isophotes show a net elongation in the EW
 direction (Fig. \ref{fig:macsj0553}). From the analysis of Chandra
 data, \citet{2012MNRAS.420.2120M} derive that the two clusters
 involved in the merger are likely to have the same mass, and from the
 separation between the two clusters, the authors conclude that the
 merger axis lies approximately in the plane of the sky.  \\ In
 Fig. \ref{fig:macsj0553} the radio emission of the cluster is
 shown. We detect a $\sim$ 4$'$ wide radio source, that is almost
 resolved out in the higher resolution image
 (Fig. \ref{fig:macsj0553}, left panel).  We classify such a source as
 radio halo.  From optical radio overlays, we identify only one
 radio-galaxy in the region of the radio halo emission (labeled by
 letter A in Fig.  \ref{fig:macsj0553}, left panel), while the radio
 emission at the centre of the cluster corresponds to the brightest
 part of the radio halo. The radio halo shows a uniform brightness,
 and its emission is elongated toward the EW direction, following the
 gas distribution, as traced by the X-ray emission. One of the X-ray
 peaks coincides with a region of higher radio brightness. The flux density,
 power, and the Largest Linear Size (LLS) of the radio halo are noted
 in Table \ref{tab:radio}.

\subsection{On the absence of radio relics in MACSJ0553.4-3342}
\label{sec:norel}
We note that in this cluster, although we are likely to witness an
energetic merger in the plane of the sky, no radio relic is
detected. According to \citet{2006MNRAS.373..881P}, the X-ray
luminosity of the cluster translates into a mass of $\sim$10$^{15}\,
M_{\odot}$. This estimate is robust against a possible uncertainty in
the mass-ratio and impact-parameter of the merger. A binary head-on
merger between clusters with this mass is expected to produce
Mpc-sized shocks in the ICM. Moreover, the geometry of the merger,
with the merger axis in the plane of the sky, should give the best
orientation to detect the relic emission \cite[e.g.][]{Vazza12}. The
two clusters are likely to have the same mass
\citep{2012MNRAS.420.2120M}. The resulting Mach number of the merger
shocks is expected to be low, since to first approximation the two
clusters will fall into each other with virial velocities. We can
roughly estimate how far a low Mach number shock could travel in the
ICM under these conditions. Following \citet{Sarazin99}, the relative
impact velocity of two subclusters with Mass $\sim$10$^{15} \,
M_{\odot} $ at the distance of 1 virial radius is $\approx$ 2400 km
s$^{-1}$. The distance between the two gas cores is now $\sim$200 kpc.
The brightest galaxies, that can be used as tracers for the Dark
Matter (DM) components, is instead $\sim$500 kpc
\citep{2012MNRAS.420.2120M}. Assuming that the infall velocity is
constant, the two gas cores would take $\sim$0.08 Gyr to achieve the
observed separation. In this time, a shock wave with a Mach number
$M\sim$2 would travel a distance of $\sim$570 kpc, hence, a
relic/shock wave should be well within the radio and X-ray fields of
view. There are however some possibilities to maintain the correlation between
relics and shock-wave: (i) the wave front is currently located along
the merging axis, close to the clusters' cores.  Such regions are
already hot, and the resulting Mach number$\leq$1. This situation
could e.g. resemble the one in MACSJ0717+3745
(\citealt{2009A&A...503..707B} but see \citealt{2009A&A...505..991V}
for a different interpretation). It is also possible that radio
  relics are actually present, but not visible because of the low
  contrast against the halo emission. Radio polarization information
  would resolve the issue. (ii) If the Mach number is low
(i. e. $\leq$ 2-3) DSA models predict a low electron-acceleration
efficiency \citep{2007MNRAS.375...77H}, although low Mach number
  shocks have been observed to produce radio emission
  \citep[e.g.][]{2005ApJ...627..733M}.  (iii) At a redshift of 0.431
  the magnetic field at the periphery of the ICM is lower than the
  equivalent Inverse Compton magnetic field ($B_{\rm {ICM}}\sim$6.6
  $\mu$G), so that IC losses dominate over synchrotron losses, and no
  relic is visible. If this is the case, we should expect to detect
  only a few relics at similar redshifts, generated by high Mach
  number shock waves. DSA is the most simple mechanism that can power
  the radio emission in relics. Other less efficient mechanisms could
  be responsible for the relic emission, as e.g. the one proposed by
  \citet{2010A&A...510A..97S}.  We note two recent works indicate that
  shock waves alone are not sufficient for powering the relic radio
  emission: \citet{2011MNRAS.417L...1R} detect two shock waves but no
  radio relic in the cluster Abell 2146. In order to explain the radio
  relic in Abell 754, \citet{2011ApJ...728...82M} claim that a direct
  acceleration of the electrons from the thermal pool to
  ultra-relativistic energies is problematic. Shock waves need to
  encounter a pre-existing population of relativistic electrons to
  produce the radio emission.

\subsection{Time-scale for particle acceleration in the radio halo}
Assuming that the merger axis is in the plane of the sky, as suggested
by X-ray observations, we have derived that the collision between the
two gas cores took place $\sim$0.08 Gyr ago.  The presence of a radio
halo in this young-merger cluster can put constraints to the
time-scale for acceleration processes. According to turbulent
re-acceleration models, the cascade time of the largest turbulent
eddies is of the order of 1 crossing time. During this time, the
diffusion and transport of the turbulent eddies may fill a large
region, and give rise to radio emission throughout the whole cluster
volume \citep{2005MNRAS.357.1313C}.  In the case of MACSJ0553.4-3342,
the crossing time for a region of 1.3 Mpc, over which the radio
emission is detected, is $\sim$ 0.5 Gyr.  If turbulence is injected
when the cores pass each other, the time-scale is not sufficient for
the turbulent process to re-accelerate the particles over such a large
and uniform volume. We conclude that either the merger is not in the
plane of the sky, contrary to what is suggested by X-ray observations,
or the turbulence started to develop well in time before the core
passage, for instance by smaller groups and sub-clamps accreting along
the merger axis \citep{2011A&A...529A..17V}. Alternatively, the radio
halo in this system cannot be explained by turbulent re-acceleration
alone.

\begin{table*} 
\caption{Radio properties }   
\label{tab:radio}      
\centering          
\begin{tabular}{|c c c c c c c|}    
\hline\hline       
Cluster       & Source  & freq &  S   & L&   LLS     & distance \\
              &         &  MHz &  mJy & 10$^{25}$  W Hz$^{-1}$ & kpc    & kpc  \\
\hline       
&&&&&&\\
MACSJ1149.5+2223 &    W-Relic  & 323   & 17$\pm$2 & 1.73 $\pm0$ 0.1 & 760  &  1140\\
MACSJ1149.5+2223 &    E-Relic  & 323   & 23$\pm$2 & 2.79 $\pm$ 0.2     & 820  &  1390\\
MACSJ1149.5+2223$^{a}$ &    Halo (?) & 323   & 29$\pm$4 & 4.77 $\pm$ 0.3     & 1300 & -\\
MACSJ1149.5+2223 &    W-Relic  & 1450  & 5.6$\pm$0.3& 0.56  $\pm$ 0.03   &  730 & 1100 \\
MACSJ1149.5+2223 &    E-Relic  & 1450  & 4.1$\pm$0.2& 0.50  $\pm$ 0.02        &  860 & 1380\\
MACSJ1149.5+2223$^{a}$  &    Halo(?)  & 1450  & $\sim$1.2$\pm$0.5& $\sim$0.19  $\pm$ 0.09      &  -& - \\
&&&&&&\\
\hline
\hline
&&&&&&\\
MACSJ1752.0+4440 &   NE-Relic  & 323    &410$\pm$33 & 19$\pm$2  &  1130  & 1300\\ 
MACSJ1752.0+4440 &   SW-Relic  & 323    &163$\pm$13 & 7.3$\pm$0.6  &  910 &  800\\ 
MACSJ1752.0+4440 &   Halo     & 323     &164$\pm$13 & 7.8$\pm$0.6    & 1650   & -\\ 
&&&&&&\\
\hline
&&&&&&\\
 MACSJ0553.4-3342 &   Halo     & 323  & 62 $\pm$ 5  & 4.4$\pm$0.4$*$   &  1300     &  -  \\
&&&&&&\\
\hline

\multicolumn{7}{l}{\scriptsize Col 1: Cluster name; Col. 2: type or
  radio source in the cluster: H$=$ radio halo, R$=$ radio
  relic;}\\ \multicolumn{7}{l}{\scriptsize Col. 3 reference frequency
  for the table entries; Col 4 and 4: Radio flux density and radio
  power;}\\ \multicolumn{7}{l}{\scriptsize Col 6: Largest Linear Scale
  of the diffuse emission; Col. 7: Projected distance from the cluster
  centre; }\\
 \multicolumn{7}{l}{\scriptsize $^{a}=$ flux density is computed above 2$\sigma$ level. Errors are
  computed assuming 10\% errors on the absolute flux density.  $*=$ we assume
  $\alpha=$1.3 for the k-correction.}
\end{tabular}
\end{table*}

\begin{table} 
\caption{Flux density of the radio sources embedded in the diffuse emission }   
\label{tab:sources}      
\centering          
\begin{tabular}{|c c c|}  
\hline\hline
Source      &   flux 323MHz  & flux 1.4 GHz     \\
& mJy & mJy\\
\hline
&&\\
MACSJ1149.5+2223 A &   5.5$\pm$0.4     & 1.51$\pm$0.09 \\
MACSJ1149.5+2223 B & 6.6$\pm$0.5  & 2.08$\pm$0.1 \\
MACSJ1149.5+2223 C &   1.8$\pm$0.1     &  -\\
MACSJ1149.5+2223 D &   1.09$\pm$0.09    &  -\\
MACSJ1149.5+2223 E &   1.3$\pm$0.1     & 0.52$\pm$0.03\\
MACSJ1149.5+2223 F &   2.9$\pm$0.2     & 0.37$\pm$0.02\\
MACSJ1149.5+2223 G &   12.5$\pm$1      & 8.2$\pm$0.5 \\

&&\\
\hline
Source      &   flux 323MHz  & flux 1.7 GHz     \\
& mJy & mJy\\
\hline
&&\\
MACSJ1752.0+4440 A &  12.2$\pm$0.9   & 3.3 $\pm$0.2   \\
MACSJ1752.0+4440 B &  6.3$\pm$0.5    & 1.5 $\pm$0.1  \\
MACSJ1752.0+4440 C &  1.8$\pm$0.1    & 3.4$\pm$0.2    \\
MACSJ1752.0+4440 D &  31.7$\pm$2.5   & 7.2$\pm$0.4     \\ 
MACSJ1752.0+4440 E &   -   &   0.37$\pm$0.02  \\ 
MACSJ1752.0+4440 F &   -   &   0.17 $\pm$0.07  \\ 
&&\\
\hline
Source      &   flux 323MHz  &      \\
& mJy & \\
\hline
&&\\
 MACSJ0553.4-3342 A &  2.0 $\pm$0.2 &  - \\
&&\\
\hline
\end{tabular}
\end{table}

\begin{figure*} 
\centering
\includegraphics[width=0.49\textwidth]{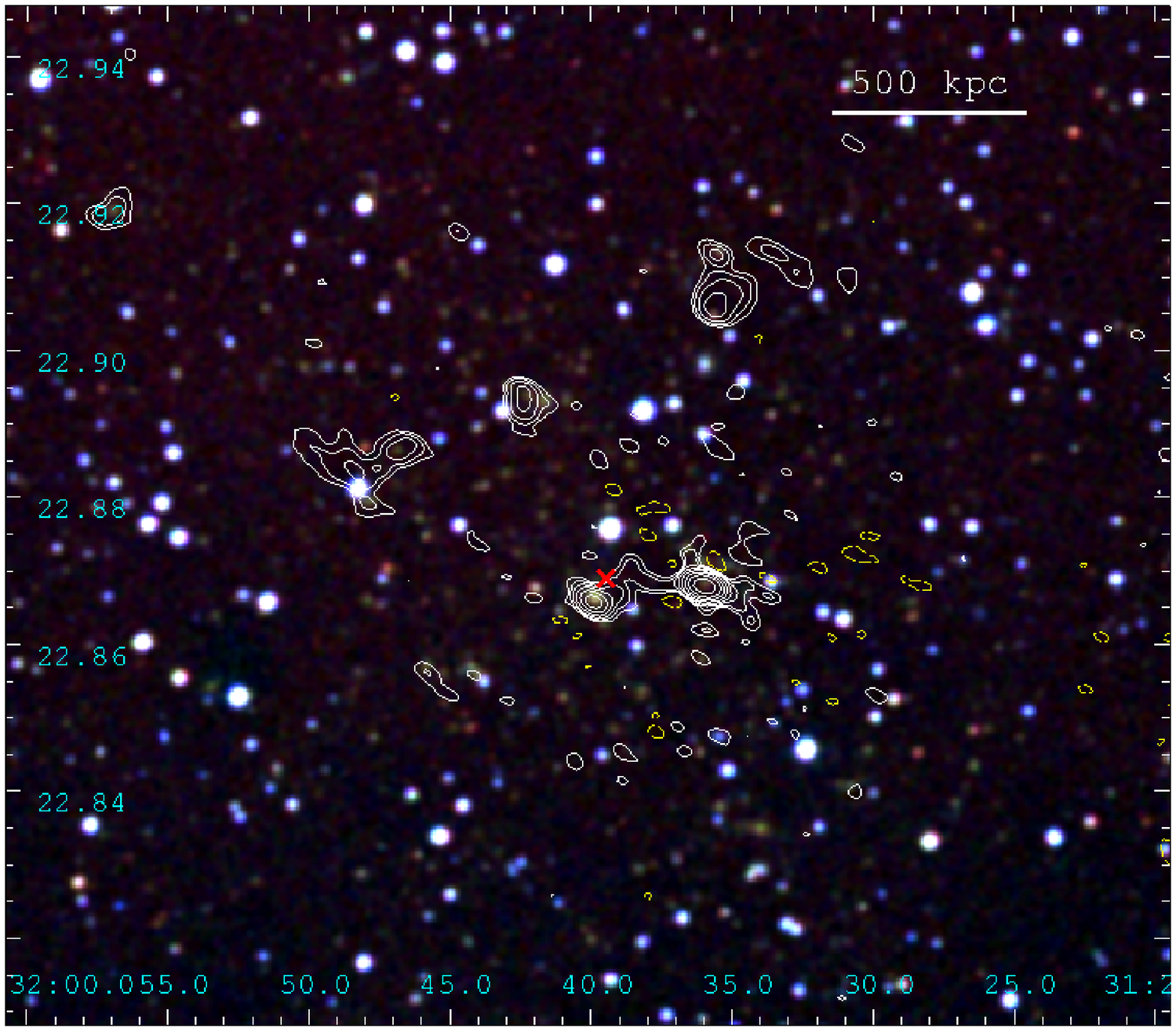}
\includegraphics[width=0.43\textwidth]{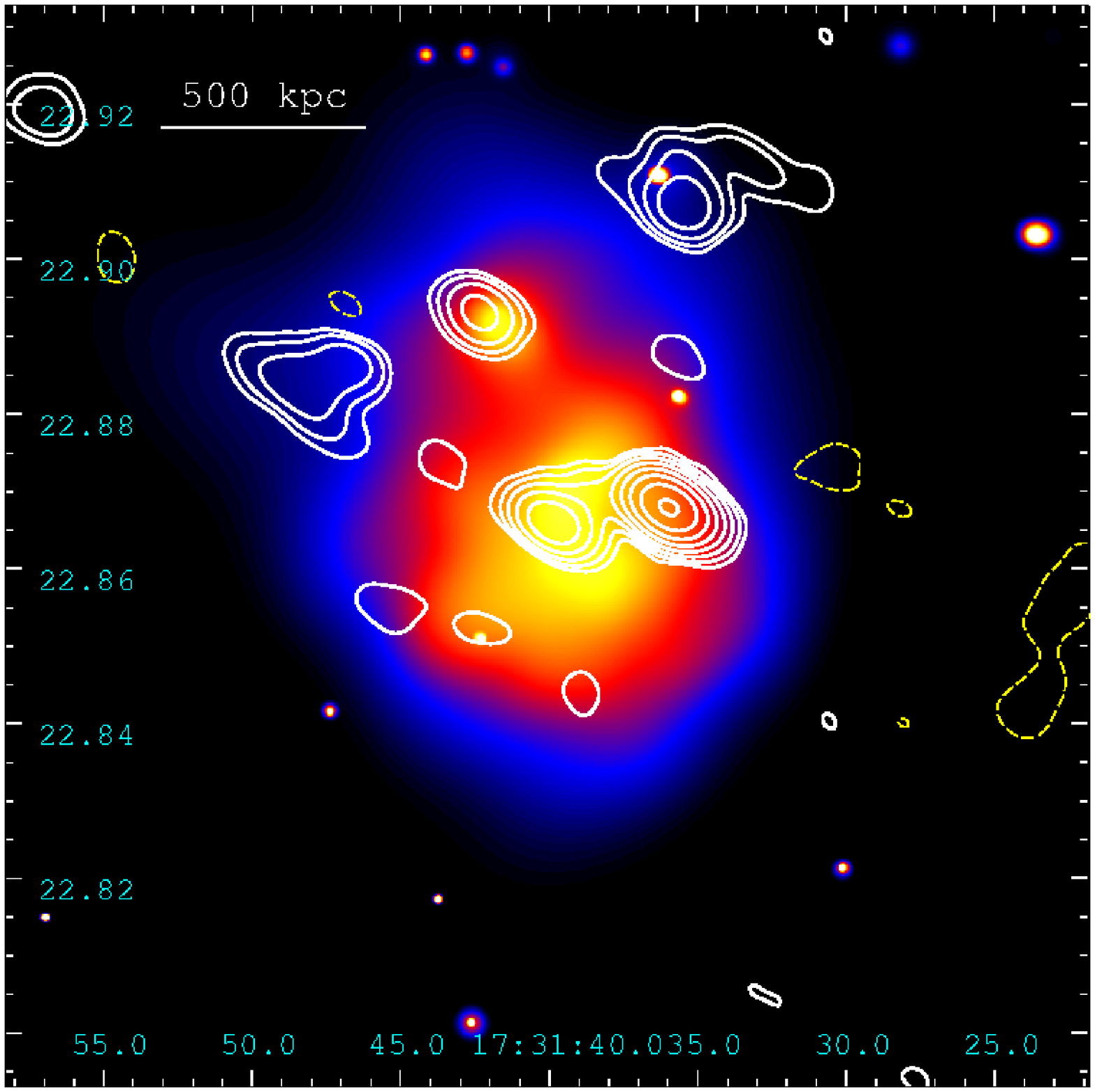}
\caption{{\bf MACSJ1731.6+2252} {\it Left:} the colour image shows the
  emission (infra-red, red and blue) from the second Digitized Sky
  Survey (DSS II) . Contours show the radio emission at 323 MHz from
  the GMRT, obtained with Robust$=$0 weighting. The restoring beam is
  10.6$'' \,\times$ 6.6$''$. The rms noise is $\sigma \sim$ 0.3
  mJy/beam. Contours start at 4$\sigma$ and are spaced by powers of
  2. The -4$\sigma$ contour is displayed with dotted yellow lines. The
  red cross at the centre marks the position of the X-ray
  centre.  {\it Right:} X-ray
  emission of the galaxy cluster as seen by {\it Chandra} ACIS-I
  detector, in the energy band 0.5-7 keV \citep{Ebeling10}. Contours
  show the radio emission at 323 MHz from the GMRT obtained with a
  Gaussian taper of the longer baselines and Robust$=$1
  weighting. The restoring beam is 26$'' \, \times$ 17$''$. The rms
  noise is $\sigma \sim$0.5 mJy/beam. Contours start at 4$\sigma$ and
  are spaced by powers of 2. The -4$\sigma$ contour is displayed with
  dotted yellow lines.}
\label{fig:macsj1731}
\end{figure*}

\subsection{MACSJ1731.6+2252}
MACSJ1731.6+2252 was classified as an active merger by
\citet{2012MNRAS.420.2120M} in a study based on a comparison of
Chandra and optical images. The authors suggest that the main
cluster, which shows no significant optical-X-ray offset, has just
undergone a merger with a much less massive subcluster.  We should be
seeing the system after the core passage.  The lack of a significant
offset between optical and X-ray components might suggest a collision
at large impact parameter, or a significant inclination of the merger
axis with respect to the plane of the sky
\citep{2012MNRAS.420.2120M}. The Chandra image and the optical DSS
image are shown in Fig. \ref{fig:macsj1731}. More details about the
X-ray properties of the cluster are listed in Table
\ref{tab:xray}.\\ The radio emission of the cluster is shown in
Fig. \ref{fig:macsj1731}.  The hint of diffuse radio emission, visible
in the NVSS image, results from the blending of the discrete radio
sources.  MACSJ1731.6+2252 is an example for a cluster that shows no
diffuse radio emission, although it is in a clear merging state. In the
re-acceleration model, this can be explained if
the merger is minor, as the optical and X-ray data suggest
\citep{2012MNRAS.420.2120M}. In this case, the energy released into the
ICM may not have been sufficient to power the radio emission at the sensitivity
level of these observations.  Observations at lower frequencies may
detect the emission from the less energetic particles.
Alternatively, not enough time may have elapsed to transfer energy from the large-scale turbulence to the relativistic particles.
According to \citet{2005MNRAS.357.1313C}, the cascade time
of the largest eddies of MHD turbulence to the acceleration scale is of the order of $\sim$
1 Gyr (see however Sec. \ref{sec:norel}).

\section{Discussion}
\label{sec:discussion}
\subsection{Shock properties from radio emission}
If relics originate from shock acceleration, the radio properties of the
observed emission reflect the properties of the shock wave.  In
particular, in the regime of DSA, the Mach number, $M$, of the shock is
related to the observed radio spectral index through \citep{1987PhR...154....1B}
\begin{equation}
\alpha_{i}=-\frac{1}{2}+\frac{M^2 +1}{M^2 -1}
\end{equation}
where $\alpha_{i}$ refers to the injection spectral index. If we
assume that particles are accelerated in regions with flatter spectra,
we get $\alpha_{i} \sim$0.6 and 0.8 for the NE and SW relics of
MACSJ1752.0+4440, respectively. These $\alpha_{i}$ correspond to
$M\sim$ 4.6 and 2.8 for the NE and SW relic respectively.  In the
cluster MACSJ1149.5+2223 the flattest values of the spectral index are
0.7 and 0.6 for the E and W relic respectively, corresponding to $M
\sim$3.3, and $M \sim$ 4.6 for the E and W relic respectively.
According to cosmological simulations, shocks with these Mach numbers
are very rare event
\citep{2011MNRAS.418..960V,2011ApJ...735...96S,scienzo}.  Such
simulations also show that the Mach number can vary across the shock
front. In these cases, the Mach numbers derived from the radio
spectral index would be biased to higher values, since the synchrotron
luminosity strongly increases with Mach number. It would not be
surprising, then, if the Mach numbers derived from the gas properties
were slightly lower than those inferred from the radio spectral
indices.

\subsection{Comparison with numerical simulations}
\begin{figure*}
\centering
\includegraphics[width=0.99\textwidth]{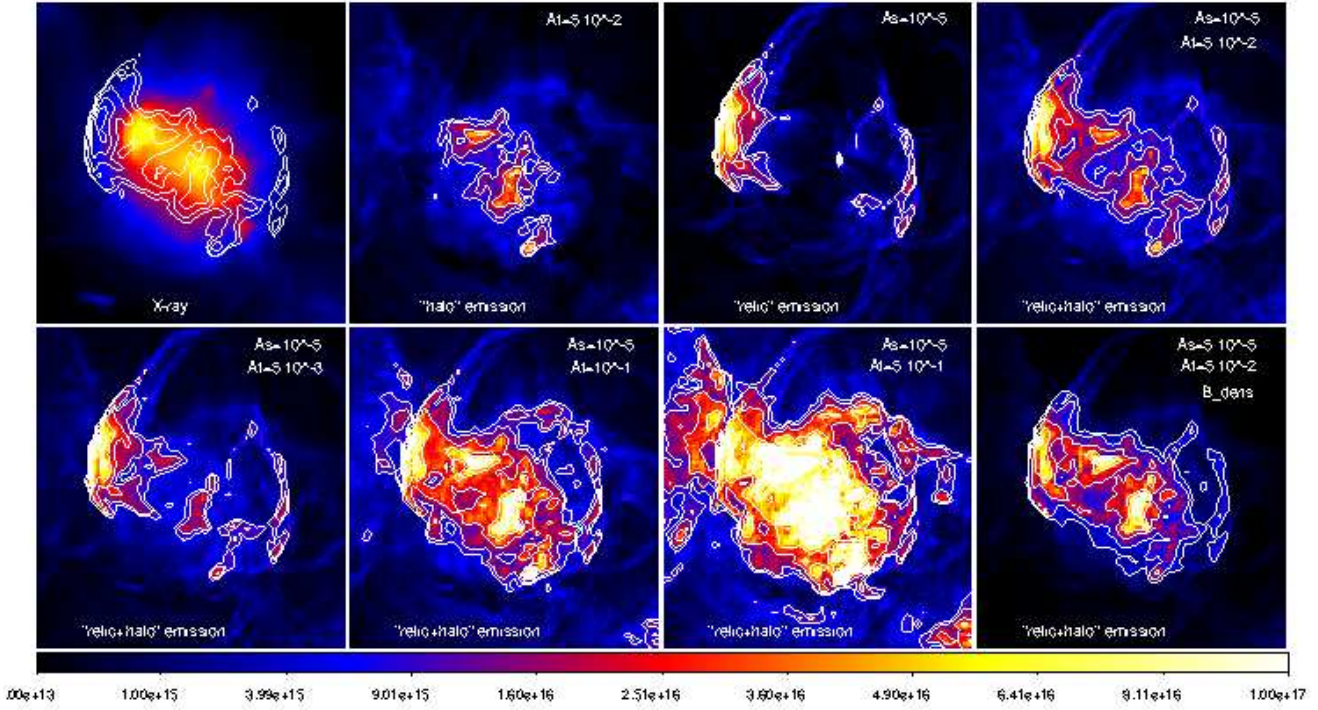}
\caption{Mock observations for a simulated cluster at $z=0.05$. Top
  left: projected X-ray flux in the [0.5-2] keV energy band
  (colours). In the remaining 3 panels of the same row: simulated
  halo, relic"emission (colours, [$10^{20} \rm erg s^{-1}$]) and
  relic+halo emission for the set of values $A_s=10^{-5}$ and
  $A_t=0.05$. The iso-contours are drawn similarly to the radio
  observation in Fig. \ref{fig:macsj1752} smoothing the numerical data
  over an equivalent beam size (we show them also in overlay with the
  X-ray colour map in the first panel, which should be compared with
  the left panel of Fig.\ref{fig:macsj1752}). The efficiencies assumed
  for the production of synchrotron at shocks and turbulence are
  printed within the panel. Lower panels: observable radio emission
  from the same region, assuming different efficiencies for shocks and
  turbulence, or a background magnetic field that scales with the gas
  density as $B \propto \rho^{0.5}$.}
\label{fig:sim1}
\end{figure*}

In this section we compare our observations with mock radio and X-ray
images obtained from cosmological simulations \citep{va10kp}.
Cosmological simulations of large volumes of the Universe routinely
provide a variety of merger configurations for evolving galaxy
clusters, which can be used to guide the physical interpretation of
observed radio and X-ray features \citep[e.g.][for recent comparison
  between cosmological simulations and observations of particular
  objects.]{2010MNRAS.401...47D,2008A&A...484..621S} We tried to find
constraints to our observations in a sample of clusters produced in
high-resolution cosmological simulations
\citep{va10kp,2011A&A...529A..17V}. These simulations were performed
using the ENZO code \citep[e.g.][and references
  therein.]{2010ApJS..186..308C} We inspected the projected X-ray
images (in the 0.5-2 keV band) of each object along many projections,
for $\sim 200$ simulated outputs in the range $0 \leq z \leq 1$.
Based on the morphology of the images, we found one case which was
very similar to the X-ray data of MACSJ1752.0+4440. We then computed
the structure of turbulent motions and shock waves in this system, and
estimated the kinetic energy flux dissipated by turbulent motions and
shock waves, studying their possible link to the observed large-scale
radio emission.  Figure \ref{fig:sim1} shows the projected X-ray
images of this cluster merger. The top left panel shows the X-ray and
radio emission of a merging cluster at $z=0$, about $\sim 0.8-1$ Gyr
after the collision of the cores. The total virial mass of the system
at this redshift is $\approx 0.65 \cdot 10^{15} \rm M_{\odot}$. In
post-processing, we measured the strength of shocks and turbulent
motions at high resolution. Thus we obtain an estimate of the power
that shocks and turbulent motions can provide to particle acceleration
at the epoch of the observation. Shock waves are detected with an
algorithm based on velocity jumps and divergence as well as the
temperature jump. With this technique we can assign a Mach-number to
each shocked cell in the simulation, and compute the kinetic energy
flux across each shocked surface. For detailed information on this
algorithm we refer the reader to \citet{va09shocks}. Turbulent motions
are extracted from the velocity field with an algorithm recently
presented in \citet{2012arXiv1202.5882V}.  This algorithm recursively
analyzes the local mean velocity field around each simulated cell, and
reconstructs the maximum correlation scale of turbulent motions across
the simulated volume.  In order to estimate the radio emission from
shock acceleration \citep[e.g.][]{2007MNRAS.375...77H} and turbulent
re-acceleration \citep[e.g.][]{2001MNRAS.320..365B}, we followed the
simple approach of producing projected maps of energy flux from
relativistic electrons accelerated by shocks and turbulence: The
kinetic flux through the shock is $\Phi_{\rm shock}=\int{\rho v_{\rm
    s}^3 dx^2}/2$ (where $\rho$ is the up-stream gas density, $v_{\rm
  s}$ is the shock velocity and $dx$ is the cell size), and $\Phi_{\rm
  turb}=\int{\rho \sigma_{\rm t}^3 dx^3}/(2 \cdot l_{\rm t})$ (where
$\sigma_{\rm t}$ is the local turbulent velocity dispersion, and
$l_{\rm t}$ is local outer scale of turbulence) for the turbulent
kinetic energy flux.\\ At a given frequency, we assume that energy is
converted from shocks with an efficiency $A_{\rm s}$ such that $P_{\rm
  relic} \sim A_{\rm s} \Phi_{\rm shock}$, and for turbulence with an
efficiency $A_{\rm t}$ such that $P_{\rm halo} \sim A_{\rm t}
\Phi_{\rm turb}$.  Various physical processes can set these conversion
factors, such as DSA, particle-wave couplings, and damping with
thermal plasma. The local magnetic field is also folded into these
conversion factors. In addition, inefficient turbulent re-acceleration
may cause a high time-variability of particle spectra
\citep[e.g.][]{2001MNRAS.320..365B}.  Here we apply these simple
recipes to estimate the morphologies and intensities of radio power
from the simulated cluster. To start with, we assumed a constant
magnetic field strength $B=1 \mu G$ across the whole simulated volume,
and tuned the efficiency for conversion of kinetic energy into
synchrotron emission at shocks, $A_{\rm s}$ to fit the observed power
and morphology of the NE relic ($\sim 2 \cdot 10^{26}$ erg s$^{-1}$
Hz$^{-1}$ at 325 MHz). The simulated NE relic (third panel in the top
row of Fig.\ref{fig:sim1}) is produced by a shock with an average Mach
number of $M \approx 4.5-5.5$, very similar to the value derived from
the radio spectral index.  The total radio power for this region is
equal to the one of the NE relic in MACSJ1752.0+4440 if we impose $A_s
\approx 10^{-5}$. For this rather strong shock, this value of $A_s$
would imply an electron to proton acceleration efficiency in the range
of $R_{\rm e/p} \sim 10^{-3}-10^{-4}$
\citep[e.g.][]{2007MNRAS.375...77H}, in line with recent numerical
results \citep[e.g.][]{2011ApJ...735...96S,2012MNRAS.420.2006N}.  We
then varied $A_t$ in the simulation to match the observed power and
morphology of the central halo in MACSJ1752.0+4440. We find that the
best match is achieved for $A_t \sim 0.05$. Typical values of the
turbulent velocity measured in this region are of the order of $\sim$
500 km s$^{-1}$ (while the bulk motions of the ICM in the "halo"
region are of the order $\sim$ 1500-2000 km s$^{-1}$).  We note that
our estimate of $A_t$ is different from the efficiency of turbulent
acceleration, $\eta$ assumed in theoretical models. Usually, the
efficiency of turbulent (re)acceleration in theoretical models is
scaled to the {\it pre-impact} kinetic energy of the two colliding
clusters \citep[e.g.][]{2003ApJ...584..190F,2005MNRAS.357.1313C},
while in our simulations we compute the turbulent kinetic energy $\sim
0.8-1 \rm Gyr$ after the collision. Our estimate of $A_t$ is
consistent with a value of $\eta \sim 10^{-4}$, in line with
theoretical works \citep[e.g.][and references
  therein]{2003ApJ...584..190F,2005MNRAS.357.1313C}.


We show in the lower panels of Fig.\ref{fig:sim1} the results obtained
adopting different values of $A_t$ (first three panels), and imposing
a background magnetic field that scales with the gas density, as $B
\propto \rho^{0.5}$, as suggested by both observations and
cosmological simulations
\citep[e.g.][]{2009MNRAS.398.1678D,2010A&A...513A..30B,2010MNRAS.408..684S,2011A&A...530A..24B,2011MNRAS.418.2234B}. This
shows the rather small range of parameters allowed by the
morphological comparison, and supports the choice of our fiducial
parameters for $A_s$ and $A_t$.  Considering that these simulations
\citep[][]{va10kp} were not produced to fit this particular
observation, it is remarkable that the final map of projected "radio"
flux density - and its relation to the underlying X-ray emission - is so
similar to the real observation of MACSJ1752.0+4440, both in produced
morphologies and total power.  It is not easy to relate the efficiency
we derive, particularly the one for turbulence, to theoretical works,
since we cannot model the small-scale coupling between turbulent MHD
waves, relativistic particles and thermal plasma. Moreover, in these
simulations we cannot follow the injection and advection of cosmic ray
particles over time, and the production of secondary electrons
\citep[see however][]{scienzo}. However, this test shows that a simple
binary merger with a mass ratio of $\sim 1/3$ can reproduce the
observation of MACSJ1752.0+4440 using a very tiny fraction ($\sim
10^{-5}$ at the relic and $\sim 0.05$ inside the halo) of the energy
dissipated at the shock and of the turbulent energy.\\

For the other clusters presented in this article, we have not found
a good match with the available dataset of simulated clusters
presented in \citet{va10kp,2011A&A...529A..17V}. Large cosmological runs, as the
one presented here, typically have a limited time sampling, and hence the chance of finding a good match for a complex merger configuration
with a high impact velocity are fairly low.

\begin{table*} 
\caption{Properties of double relics from the literature an this work.}   
\label{tab:relics}      
\centering          
\begin{threeparttable}

\begin{tabular}{|c c c c c c c c c c c |} 

\hline
\hline
name  &   $z$  &     freq &   flux & err flux    &   alpha &  $L_{1.4 \; \rm{GHz}}$ & LLS &  dist &  Lx [0.1-2.4]  &   ref \\
      &      &   GHz    &  mJy   & mJy        &         &  W Hz$^{-1}$        & kpc  & kpc    & 10$^{44}$ erg s$^{-1}$  &   \\
&&&&&&&&&&\\
\hline
A3365 E    &  0.0926  &  1.4 &  42    &   3   &   -     & 8.7e+23   &560  &  1079 &  0.859 & vW11a\\	
A3365 W    &  0.0926  &  1.4 &   5    &   0.5 &   -     & 1.0e+23   &235  &   777 &  0.859 & vW11a\\
ZwCl0008.8+5215 E &  0.103   &  1.4 &  37    &   3   &	  1.6	& 1.0e+24   &1400 &   798 &  0.5   &  vW11b\\
ZwCl0008.8+5215 W &  0.103   &  1.4 &   9    &	  1   &	  1.5	& 2.4e+23   &290 &   808 &  0.5   &  vW11b\\
A1240 N	   &  0.1590  &  1.4 &   6.0  &   0.2 &	  1.2 	& 4.0e+23   &640 &   700 &  1.0   &  B09\\
A1240 S	   &  0.1590  &  1.4 &  10.2  &   0.4 &	  1.3	& 7.0e+23   &1250 &  1100 &  1.0	&  B09\\
A2345 E	   &   0.1765 &  1.4 &  29.0  &   0.4 &	  1.3   & 2.5e+24   &1500 &   890 &  5.3	&  B09\\
A2345 W    &  0.1765  &  1.4 &  30.0  &   0.5 &   1.5   & 2.7e+24   &1150 &  1000 &  5.3	&  B09\\
RXCJ1314.4-2515 E &  0.2439  &  0.6 &  28.0  &   1   &   1.4	& 1.6e+24   &910 &   577 &  10.9	&  V,F\\
RXCJ1314.4-2515 W &  0.2439  &  0.6 &   65   &	  3   &	  1.4	& 3.7e+24  &910 &   937 &  10.9  &  V,F\\
MACSJ1149.5+2223 E & 0.54    &  0.3 &   29   &	  3   &	  1.2   & 5.4e+24   &820 &  1390 &  14.0  &  *\\
MACSJ1149.5+2223 W & 0.54    &  0.3 &   29   &	  3   &   0.8   & 9.0e+24   &760 &  1140 &  14.0  &  * \\
MACSJ1752.0+4440 NE & 0.366   &  1.7 &   55   &	  3   &	  1.21	& 3.2e+25   &1340 &  1130 &  8.2   & vW12, * \\
MACSJ1752.0+4440 SW & 0.366   &  1.7 &   26   &	  1   &	  1.13	& 1.4e+25   &860 &   910 &  8.2   & vW12, *\\
0217+70 E  &  0.065   &  1.4 &   -    &   -   &   1.76  & -   &-   &  1000 &  -     & Br\\ 
0217+70 W  &  0.065   &  1.4 &   -    &   -   &   1.46  & -   &-   &  1000 &  -     & Br \\
A3376 E    &  0.046   &  1.4 &  166   &   8.3*&    -    & 7.9e+23   &820  & $\sim$1300 &  2.12  & Ba06\\
A3376 W    &  0.046   &  1.4 &  133   &   6.7*&    -    & 6.3e+23   &980  & $\sim$1300 &  2.12  & Ba06\\
A3667 E    &  0.056   &  1.4 &  300   &   0.2 &    -    & 2.1e+24   & 1420  &  1420   & 9.3    & R,JH\\
A3667 W    &  0.056   &  1.4 &  3700  &   300 &    2    & 2.6e+25   & 1800   & 1830   & 9.3    & R, JH\\
CIZAJ2242.8+5301 N & 0.192   &  1.4 &   -    &  -   &   1.08   & -   &1700 & 1030  & 6.8    & vW10 \\
CIZAJ2242.8+5301 S & 0.192   &  1.4 &   -    &  -   &     -    & -   &1450 & 1520  & 6.8    & vW10 \\
ZwCl2341.1+0000 N &0.27  &  0.6 &   14   &  3   &   0.5    & 3.8e+24   &250  & 1180  &  -     & G/vW09\\
ZwCl2341.1+0000 S &0.27  &  0.6 &   37   &  13  &   0.8    & 1.7e+24   &1200 & 760   &  -     & G/vW09\\
PLCKG287.0+32.9 N     &0.39  & 0.15 & 550    &  50  & 1.26     & 1.8e+25  & 1934 & 1580  & 17.2   & Ba11\\
PLCKG287.0+32.9 S     &0.39  & 0.15 & 780    &  50  & 1.54    &  1.5e+25  & 1628 & 3000  & 17.2   & Ba11\\
\hline

\multicolumn{11}{l} {\scriptsize Col. 1: Relic name, Col 2: redshift; Col. 3; frequency at which the flux density is measured}\\
\multicolumn{11}{l} {\scriptsize Col. 4 radio flux; Col. 5: error on the measured flux; Col. 6: spectral index; }\\
\multicolumn{11}{l} {\scriptsize Col. 7 Radio power at 1.4 GHz. When $\alpha$ was not available we have assumed $\alpha$=1.2}\\
\multicolumn{11}{l} {\scriptsize Col. 8: Largest Linear Size of the radio emission; Col 9: Distance from the cluster centre}\\\multicolumn{11}{l} {\scriptsize Col. 10: X -ray luminosity in the 0.1- 2.4 keV band; Col. 11: reference for the radio emission}

\end{tabular}

\begin{tablenotes}

 \item[*] This work,\item[B09]   \citet{2009A&A...494..429B},
 \item[vW11a] \citep{2011A&A...533A..35V},
 \item[vW11b] \citep{2011A&A...528A..38V},
 \item[vW10]  \citep{2010Sci...330..347V},
 \item[vW09]   \citep{2009A&A...506.1083V},
 \item[V]   \citep{2007A&A...463..937V},
 \item[F]   \citep{2005A&A...444..157F},
 \item[vW12]  \citep{2012arXiv1206.2294V}
 \item[Br]   \citep{2011ApJ...727L..25B},
  \item[Ba06]  \citep{2006Sci...314..791B},
  \item[R]  \citep{1997MNRAS.290..577R},
  \item[JH]  \citep{2004rcfg.proc...51J},
  \item[vW10] \citep{2010Sci...330..347V},
  \item[G]   \citep{2010A&A...511L...5G},
 \item[vW09] \citep{2009A&A...506.1083V},
    \item[Ba11] \citep{2011ApJ...736L...8B}.
\end{tablenotes}
\end{threeparttable}
\end{table*}
\begin{figure}
\centering
\includegraphics[width=0.5\textwidth]{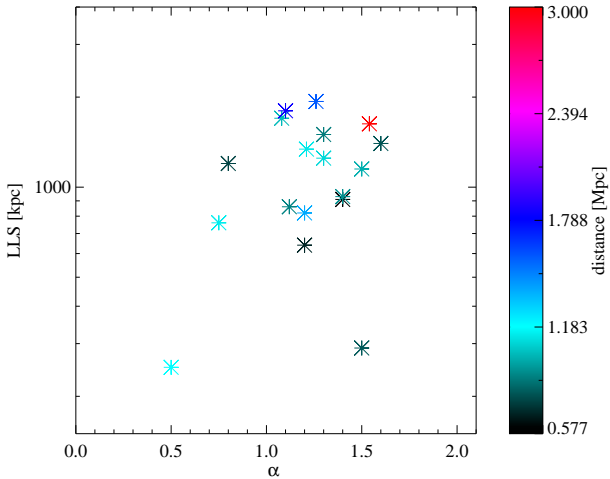}
\caption{LLS-$\alpha$ correlation for relics in double-relics
  systems. In the on-line version of the paper the distance of the
  relics to the cluster centre is colour-coded.}
\label{fig:relics_llsalphad}
\end{figure}

\begin{figure*}
\centering
\includegraphics[width=0.48\textwidth]{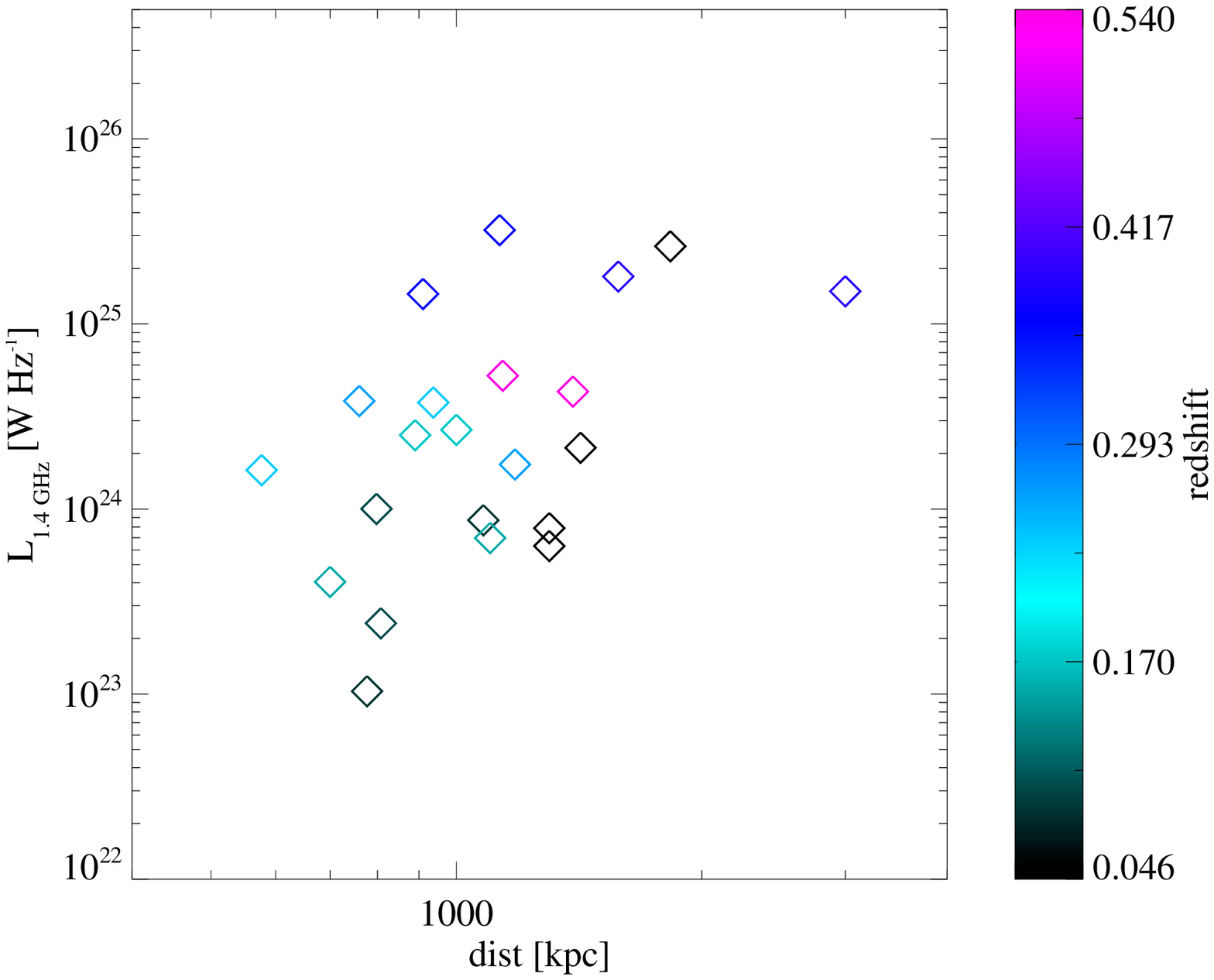}
\includegraphics[width=0.48\textwidth]{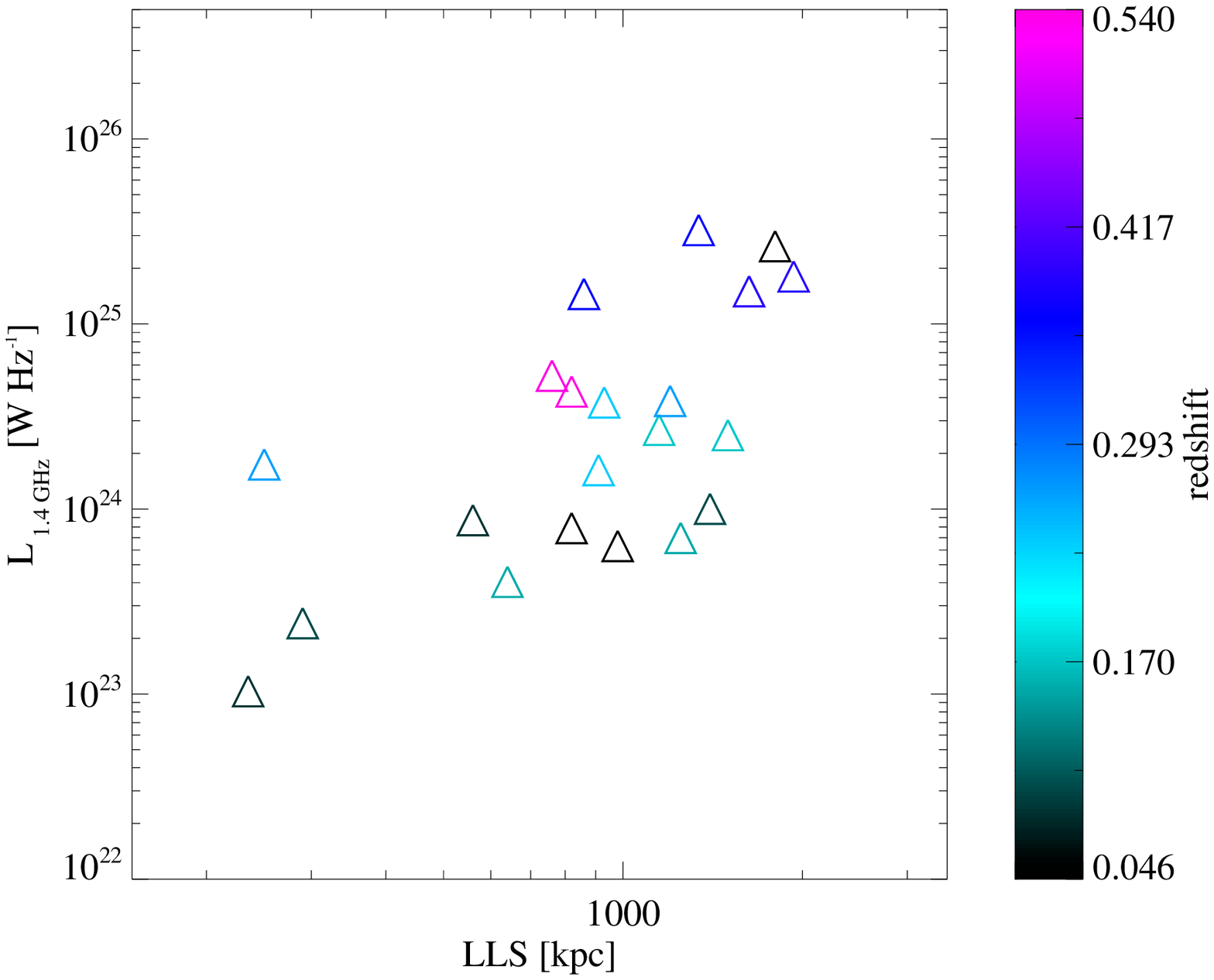}
\caption{Radio power of double relics vs distance from the cluster
  centre (left) and versus the relic LLS (right). In the on-line
  version of the paper the redshift of the hosting cluster is
  colour-coded.}
\label{fig:relics_corr}
\end{figure*}
\section{New correlations}
\label{sec:correlations}
\subsection{Double relics}
Collecting data from the literature, \citet{2009A&A...508...75V} have
found that the relic LLS correlates with the projected distance from
the cluster centre (i.e. larger relics are mostly located at larger
distances from the cluster centre), and that the spectral index
anti-correlates with the relic LLS (i.e. smaller relics use to have
steeper spectral indices). These correlations can be explained by the
fact that the larger shock waves occur mainly in lower-density and
lower-temperature regions, and have hence larger Mach numbers
\citep{sk08,va09shocks}. However, correlations that include measures
of length are affected by projection effects. As mentioned my
\citet{2009A&A...508...75V} another problem with their analysis is
that possible "radio phoenices" were also included (located close the
the cluster centre and having very steep radio spectra).  When two
relics are observed in the same cluster, the most plausible scenario,
in the framework of shock acceleration models, is that the merger is
seen in the plane of the sky. Hence, projection effects should have
the minimum impact on the measured lengths. In recent years the number
of clusters with double relics has increased significantly. The data
from the literature plus those presented in this work allow us now to
look at the same correlations using only double relics clusters. This
should reduce as much as possible the contamination by projection
effects. Data have been corrected for the cosmology adopted in this
paper, and the radio power has been computed at 1.4 GHz. The results
are listed in Table \ref{tab:relics}. In cases where the spectral
index was not available in the literature, we assumed a spectral
index $\alpha=1.2$ in order to compute the radio power. These relics
have not been considered for the spectral index correlations
investigated below.\\ In Fig. \ref{fig:relics_llsalphad} we show the
$\alpha-\rm {LLS}$ trend for double relics only. The distance is
colour-coded, as in \citet{2009A&A...508...75V}. In Fig.
\ref{fig:relics_corr} the radio-power at 1.4 GHz is plotted versus the
projected distance from the cluster centre, and versus the relic
LLS. In these two plots the redshift of the host cluster is
colour-coded.  To investigate the existence of possible correlations
between the plotted quantities, we use the Spearman's Rank Correlation
Coefficient (S-rank).  The null hypotheses we are testing is the
following: ``$H_0$: there is no association between var$_a$ and
var$_b$.''  We investigate separately a possible correlation between:
\begin{itemize}
\item{$L_{1.4 \rm{GHz}}$ and distance from the cluster centre}\\
\item{LLS and distance from the cluster centre}\\
\item{LLS and $L_{1.4 \rm{GHz}}$}\\
\item{LLS and $\alpha$}\\
\item{$L_{1.4 \rm{GHz}}$ and $z$}
\end{itemize}
We have computed the S-rank and we have tested for correlations at the
5\% level of significance (2-tail test). We find that the null
hypotheses is accepted for $L_{1.4 \rm{GHz}}$-dist, LLS-dist,
LLS-$\alpha$, and $L_{1.4 \rm{GHz}}-z$, meaning that no correlation is
significantly detected in these clusters at the 5\% significance
level. We note that if we apply the 1-tail test, the correlations
$L_{1.4 \rm{GHz}}$-dist, LLS-dist, and $L_{1.4 \rm{GHz}}-z$ are found
to be significant. Such a test is not
  accurate enough to establish a robust statistical correlation, but we
  argue that these correlations should be investigated using
  volume-complete samples. The null hypothesis is instead rejected
for the correlations LLS-$L_{1.4 \rm{GHz}}$, meaning that the two
variables are correlated at 5\% significance level. Relics with high
radio power tend to have larger linear sizes.\\


\begin{figure*}
\centering
\includegraphics[width=0.48\textwidth]{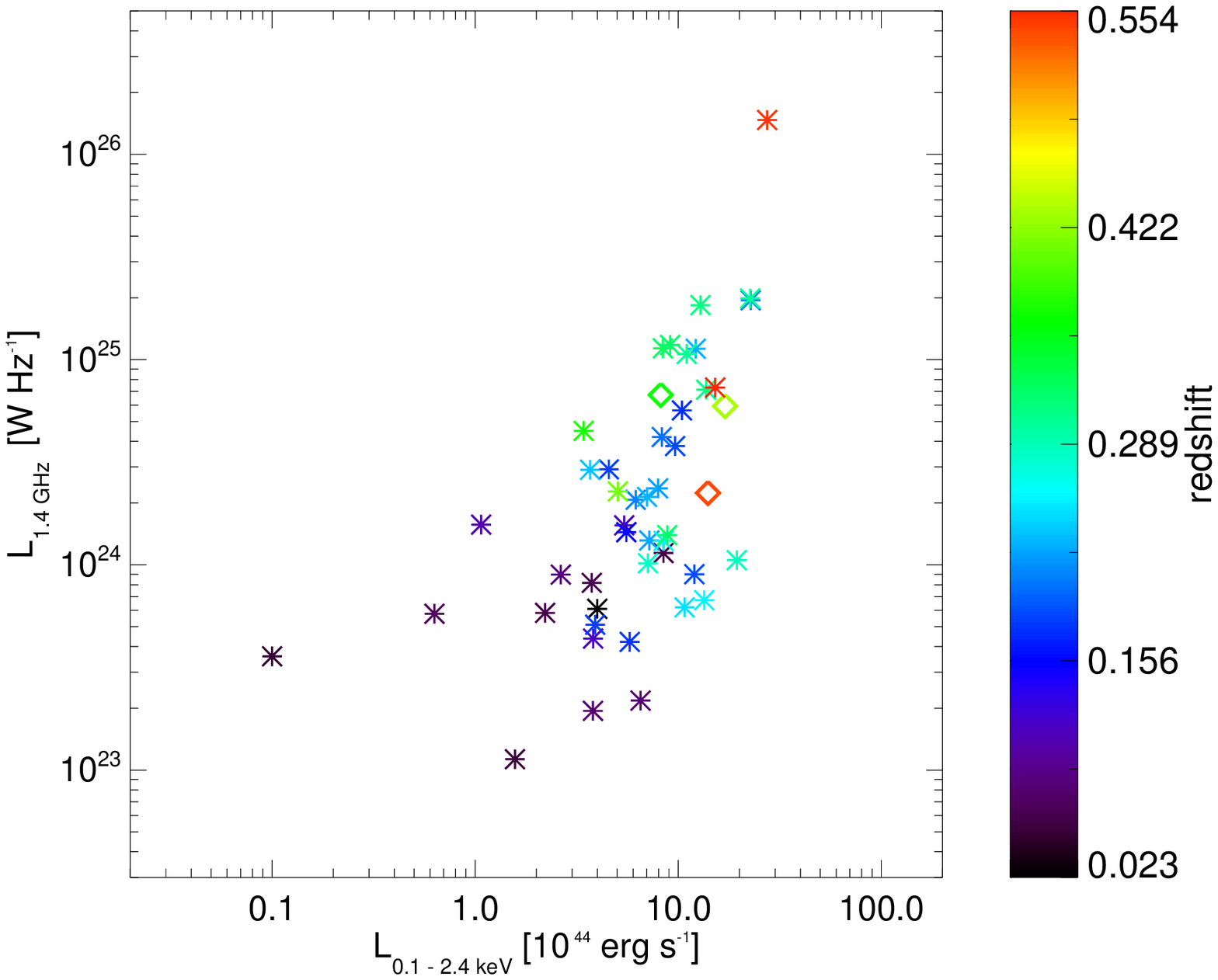}
\includegraphics[width=0.48\textwidth]{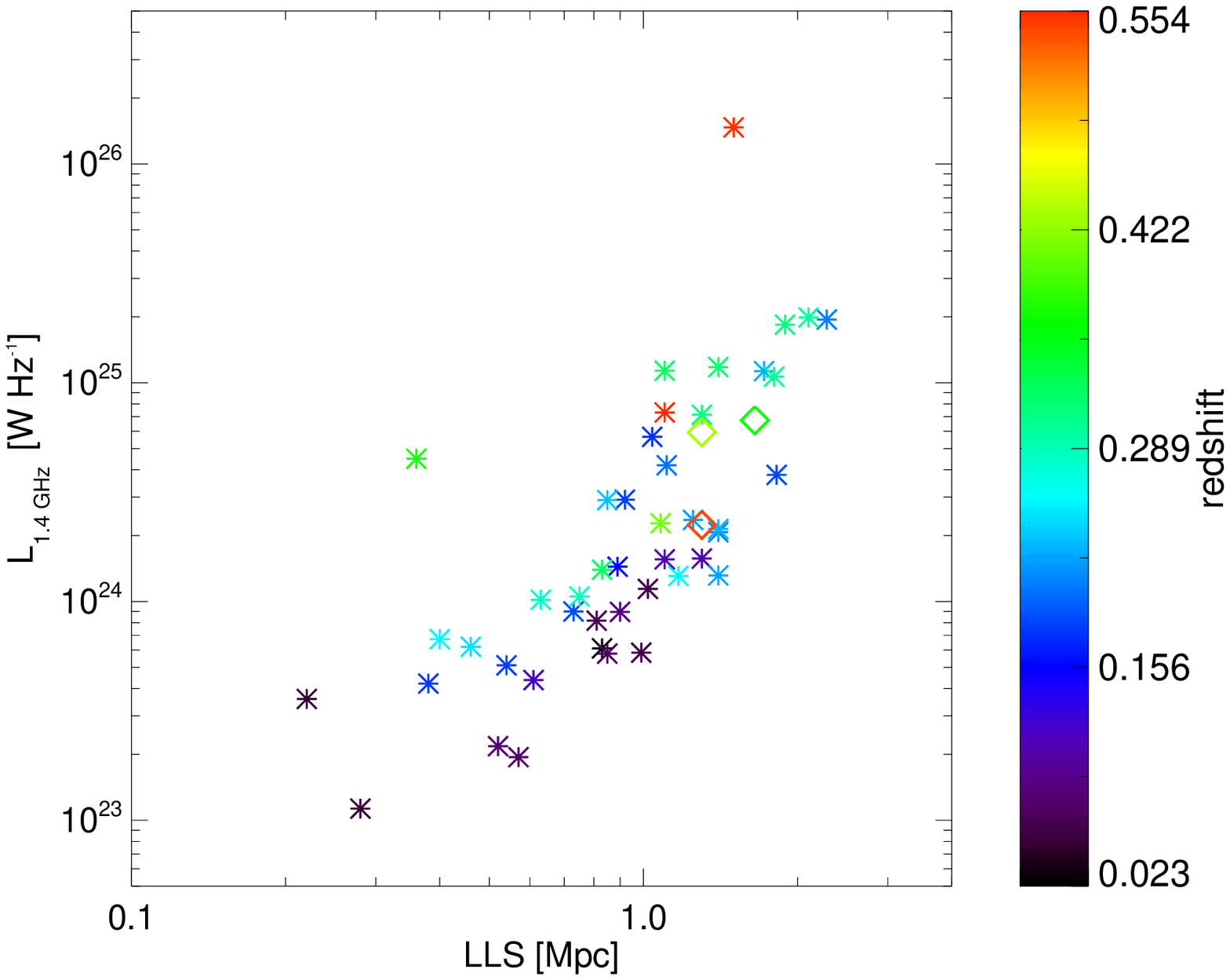}
\caption{Left panel: Correlations between the radio halo power at 1.4
  GHz ($L_{\rm 1.4 Ghz}$) and the cluster X-ray Luminosity computed in
  the 0.1-2.4 keV energy band ($L_x$). Right panel: Correlations
  between the radio halo power at 1.4 GHz ($L_{\rm 1.4 Ghz}$) and the
  Largest Linear Scale of the radio emission. In the on-line version
  of the paper the redshift is colour-coded. The newly discovered
  radio halos are marked with diamonds.}
\label{fig:halos}
\end{figure*}

\subsection{Radio halos}
\citet{Giovannini09} have compiled a list of the radio halos at
$z<$ 0.4 known to that date. In this work, we have more than doubled
the radio halos in clusters at $z>$ 0.4. Although the sample we have
presented focuses on observations of the most promising candidates,
and is not a statistical sample, it is interesting to compare the
properties of higher and lower redshift radio halos. We have taken all
the clusters in \citet{Giovannini09} plus Abell 399
\citep{2010A&A...509A..86M} , Abell 746, RXC J0107.8+5408
\citep{2011A&A...533A..35V}, RXC J2003.5-2323,
\citep{2009A&A...505...45G}, CL0016+16 \citep{2000NewA....5..335G},
MACSJ0717+3745 \citep{2009A&A...503..707B,2009A&A...505..991V}, Abell
781 (\citealt{2011A&A...529A..69G}, but see
\citealt{2011MNRAS.414L..65V} for a different interpretation), Abell
1689 \citep{2011A&A...535A..82V}, RXCJ1514.9-1523
\citep{2011A&A...534A..57G}, Abell 1682 \citep{2011JApA...32..501V},
Abell 523 \citep{2011A&A...530L...5G}, and 0217+70
\citep{2011ApJ...727L..25B}, plus those discovered in this work. The
radio power has been computed at 1.4 GHz using the spectral indices
available in the literature and assuming $\alpha=1.3$ otherwise. In
Fig. \ref{fig:halos} the $L_{1.4 \rm{GHz}}-L_x$ correlation and the
LLS-$L_{1.4 \rm{GHz}}$ correlations are shown. The $L_{1.4
  \rm{GHz}}-L_x$ and LLS-$L_{1.4 \rm{GHz}}$ correlations are well
known \citep[e.g.][and ref therein]{Giovannini09}.  


To investigate the apparent correlations between the LLS and the
cluster redshift and between the $L_{1.4 \rm{GHz}}$ and the cluster
redshift, we have computed the S-rank. Using a 2-tail test the
correlations are confirmed with a 1\% level of significance, but we
are aware that several observational biases have to be considered:
high-$z$ samples, as MACS, cover a volume that is larger than low-$z$
ones, making it easier to pick-up the most extreme systems, and less
powerful objects are more difficult to detect at high $z$.  On the
other hand, extended radio halos at low $z$ could be resolved out by
interferometers (see Fig. 15 in \citealt{Giovannini09}).  Hence,
complete statistical samples covering similar volumes, and free from
the above mentioned biases should be used to test the correlation. If
confirmed, this correlation would further support the radio-halo
merging paradigm. In fact, major mergers between massive clusters are
expected to take place in the redshift range $z>0.3-0.4$. The power
of radio halos could hence trace the quantity of energy injected in
the ICM during the cosmic time.  The dependence of the halo's LLS from
the cluster redshift is not trivial to explain. As the redshift
increases, also the IC equivalent field increases, so that one should
expect higher $z$ objects to be more dominated by IC emission compared
to the local ones. Hence, this correlation might hint towards a mechanism that is more efficient at the cluster outskirts at higher
$z$. However, complete statistical samples covering
similar volumes and free of biases are needed to draw any firm conclusions on this.

\section{Conclusions}
\label{sec:conclusions}
We have presented new radio observations of massive galaxy clusters
located at high redshift ($z>0.3$). Our GMRT observations combined
with WSRT and archival VLA observations have lead to the discovery of
two radio halos, 2 double-relic systems and a candidate radio halo.
Our results can be summarized as follows:
\begin{itemize}
\item{Two radio relics and a candidate radio halo are detected in the
  cluster MACSJ1149.5+2223, at $z=0.54$. This is the most distant cluster
  with double relics detected so far, and it shows peculiar features that
  have never been observed in other galaxy clusters. The relics have a
  total extent of $\sim$ 800 kpc and are located at 1.1 and 1.4 Mpc
  from the cluster centre, where the X-ray brightness drops down. The
  position of the relics is peculiar, and misaligned with the cluster
  merger axis. We argue that this peculiar feature could be a sign of
  additional merging in this complex system. In the framework
  of DSA, the integrated spectral indices of the relics indicate
  Mach numbers $M\sim 3$ and 4.6. The relics show an average
  polarization of 5\%, with polarization values up to 30\%. The
  polarization vectors suggest a magnetic field roughly oriented along
  the relics' main axes. The candidate radio halo has a total extent of
  $\sim$1.3 Mpc, and a radio flux density of $\sim$29 mJy at 323 MHz. The
  radio halo is barely visible in the VLA image. The spectral
  index $\alpha$ is larger than 2, making it the radio halo with the steepest
  spectrum known so far.}

\item{In the cluster MACSJ1752.0+4440 ($z=$ 0.366) two radio relics
  and a radio halo are detected at 323 MHz. Their existence was
  already suggested by \citet{2003MNRAS.339..913E} and confirmed by
  WSRT observations \citep{2012arXiv1206.2294V}. The relics have a
  total extent of 1.1 and 0.9 Mpc, and are symmetric about the cluster
  centre. The relics in this cluster are a text-book example for shock
  acceleration. The spectral index trends derived between 1.7 GHz and
  323 MHz show a clear steepening toward the cluster centre, possibly
  tracing the particle aging as we move far from the shock front.
  Assuming that the flattest spectral indices are representative of
  the injected spectrum of the particles, we derive $M \sim$ 4.6 and
  2.8, in agreement with Mach numbers expected for merging shocks in
  the outskirts of galaxy clusters. The relics are polarized on
  average at a 20\% level, and show stronger polarization - up to 40\%
  - in the outer parts. The polarization vectors indicate an ordered
  magnetic field mainly aligned with the relic main axes, in line with
  model predictions. The radio halo has an irregular morphology. Its
  emission is elongated along the NE-SW direction, connected with the
  relics' emission. The brightest radio halo regions do not correspond
  to the brightest X-ray regions of the cluster. The spectral index of
  the radio halo between 323 MHz and 1.7 GHz is $\alpha=1.33 \pm$
  0.07.}

\item{A radio halo is discovered in the cluster MACSJ0553.4-3342, at
  $z=0.431$. Its total extent is $\sim$ 1.3 Mpc. The halo is
  elongated in the EW direction, along the merger axes. The merger
  between two sub-clusters is likely to lie close to the plane of the sky, and the
  two cores have just passed each other. Given the small separation
  between the two X-ray cores, we derive that they passed each other
  $\sim$ 0.08 Gyr ago. If turbulent (re)acceleration is responsible
  for the halo emission, then turbulent motions must have developed
  well in time before the impact of the two cores.  While the merger is violent enough to power a radio halo, we note the
  absence of radio relics in this systems and speculate about the reasons. }

\item{From a set of high-resolution cosmological simulations we have
  picked a cluster whose projected X-ray surface brightness and
  morphology is similar to MACSJ1752.0+4440. We have assumed that the
  energy is converted from shocks to radio emission with an efficiency
  $A_{\rm S}$, defined such that the radio power, $P_{\rm relic}$ is
  equal to $A_{\rm S} \cdot \Phi_{\rm shock}$, and for turbulence with
  an efficiency $A_{\rm t}$ such that the halo power is $P_{\rm halo}
  \sim A_{\rm t} \Phi_{\rm turb}$, where $\Phi_{\rm shock}$ and
  $\Phi_{\rm turb}$ is the energy flux from relativistic electrons
  accelerated by shocks and turbulence, respectively. Adopting simple
  recipes, we are able to reproduce both the radio power and the
  observed morphology of the relic emission by assuming $A_s \approx
  10^{-5}$. In order to match the power and morphology of the radio
  halo we require instead $A_s \approx 0.05$.  This conversion
    factor corresponds to an efficiency of turbulent acceleration of
    $\eta \sim 10^{-4}$. In both cases only a very tiny fraction of
    the energy supplied by the merger is required to explain the
    observed radio emission.}

\item{We have investigated the LLS-$\alpha$, distance-$\alpha$,
  $L_{1.4 \rm{GHz}}$- distance, $LLS-$distance, and $L_{1.4
    \rm{GHz}}-z$ correlations for radio relics. We have considered
  only double-relics systems to minimize projection effects. We find
  no statistical evidence for such correlation at 5\% level of
  significance (2-tail test).  We have found a correlation between
  double relics LLS and the relic radio power (LLS-$L_{1.4 \rm{GHz}}$
  correlation). We note that the correlations $L_{1.4 \rm{GHz}}$-
  distance, $LLS-$distance, and $L_{1.4 \rm{GHz}}-z$ would be accepted
  if we applied a 1-tail test at 5\% level of significance. We do not
  consider here such test to be accurate enough to probe a robust
  statistical correlation, but we argue that complete samples of
  double-relic clusters are needed to understand whether these
  correlations are present or not.}


\item {Collecting published data on radio halos, we find that the power of radio halos correlates with
  the redshift of the host cluster ($L_{1.4 \rm{GHz}}-z$ correlation),
  Since the most powerful mergers are expected at $z>0.3-0.4$, this
  correlation indicates a link between the energy injected in the ICM
  and the particle acceleration efficiency and/or magnetic field
  amplification. The LLS of the radio halo is also found to correlate
  with the cluster's redshift (LLS-$z$ correlation). This correlation
  is not easily explained since ultra-relativistic particles lose more
  energy through the IC mechanism as the redshift increases. Although
  several observational biases can affect such correlations, we argue
  that, if confirmed, they would provide some clues about the
  redshift-dependence of the particle acceleration mechanism.}

\end{itemize}

\bigskip
{\bf Acknowledgments} We thank the referee, Shea Brown, for his useful
comments on the manuscript. AB, MB, FV, MH, and UK acknowledge support by the
research group FOR 1254 funded by the Deutsche Forschungsgemeinschaft:
``Magnetization of interstellar and intergalactic media: the prospect
of low frequency radio observations''. We thank the staff of the GMRT
that made these observations possible. GMRT is run by the National
Centre for Radio Astrophysics of the Tata Institute of Fundamental
Research. This research has made use of the NASA/IPAC Extragalactic
Data Base (NED) which is operated by the JPL, California institute of
Technology, under contract with the National Aeronautics and Space
Administration. We thank G. Macario, G. Brunetti, and R. Cassano for
useful discussions. FV acknowledges the usage of computational time
  under the INAF-CINECA agreement, and acknowledges C. Gheller for
  fruitful collaboration in the production of the simulations. HE
acknowledges financial support from SAO grants GO1-12153X.

\bibliographystyle{mn2e}
\bibliography{master}


\end{document}